%% file: les.tex
\input djnlx

\twelvepoint
\doublespace

\font\smcaps=cmcsc10 at 8pt

\title

{ Large eddy simulation of two-dimensional isotropic turbulence}

\author
Semion Sukoriansky,{$^1$}\vfootnote {$^1$}{Department of Mechanical Engineering, Ben Gurion University of the Negev, and the Perlstone Center for Aeronautical Engineering Studies, Beer Sheva 84105, Israel.\hfill} Alexei Chekhlov,{$^2$}\vfootnote {$^2$}{Applied and Computational Mathematics, Princeton University, Princeton, NJ 08544.\hfill} Boris Galperin,{$^3$}\vfootnote {$^3$}{Department of Marine Science, University of South Florida, St. Petersburg, FL 33701.\hfill} and Steven A. Orszag{$^2$} 
\affil

\abstract

Large eddy simulation (LES) of forced, homogeneous, isotropic,
two-dimensional (2D) turbulence in the energy transfer subrange is the
subject of this paper.  A difficulty specific to this LES and its
subgrid scale (SGS) representation is in that the energy source
resides in high wave number modes excluded in simulations.  Therefore,
the SGS scheme in this case should assume the function of the energy
source.  In addition, the controversial requirements to ensure direct 
enstrophy transfer and inverse energy transfer make the conventional 
scheme of positive and dissipative eddy viscosity inapplicable to 2D 
turbulence.  It is shown that these requirements
can be reconciled by utilizing a two-parametric viscosity introduced
by Kraichnan (1976) that accounts for the energy and enstrophy
exchange between the resolved and subgrid scale modes in a way
consistent with the dynamics of 2D turbulence; it is negative on large
scales, positive on small scales and complies with the basic
conservation laws for energy and enstrophy.  Different implementations
of the two-parametric viscosity for LES of 2D turbulence were
considered.  It was found that if kept constant, this viscosity
results in unstable numerical scheme.  Therefore, another scheme was
advanced in which the two-parametric viscosity depends on the flow
field.  In addition, to extend simulations beyond the limits imposed
by the finiteness of computational domain, a large scale drag was
introduced.  The resulting LES exhibited remarkable and fast
convergence to the solution obtained in the preceding direct numerical
simulations (DNS) by Chekhlov et al. (1994) while the flow parameters
were in good agreement with their DNS counterparts.  Also, good
agreement with the Kolmogorov theory was found.  This LES could be
continued virtually indefinitely.  Then, a simplified SGS
representation was designed, referred to as the stabilized negative
viscosity (SNV) representation, which was based on two algebraic terms
only, negative Laplacian and positive biharmonic ones.  It was found
that the SNV scheme performed in a fashion very similar to the full
equation and it was argued that this scheme and its derivatives should
be applied for SGS representation in LES of quasi-2D flows.

\footnote {} {{\bf {SUBMITTED TO Journal of Scientific Computing}} 
March 1995}
\endtitlepage

\body

\def \w {\omega}

\def \om {\Omega}
\def \k {{\bf k}}

\def \e {\epsilon}
\def \D {\delta}
\def \DC {\Delta}

\def \eps {\overline {\epsilon}}
\def \epsl {\overline {\epsilon}_{\hbox{\smcaps {LES}}}}
\def \etal {\overline {\eta}_{\hbox{\smcaps {LES}}}}
\def \omb {\overline {\Omega}}
\def \Pb {\overline {P}} 
\def \Eb {\overline {E}} 
\def \L {\Lambda_0}

\def \p {{\bf p}}
\def \q {{\bf q}}

\def \hk {\hat k}

\def \qs {q^2}
\def \qsi { {1 \over \qs} }
\def \qksi { {1 \over {\vert \q - \k \vert}^2 }}
\def \hq {\hat q}

\pageno=2

\head {1. Introduction}

\par  Homogeneous and isotropic turbulence has been a traditional
idealization of real turbulent flows which are usually neither
homogeneous nor isotropic.  However, this idealization provided
wealth of information on the physics of turbulence and it still
remains one of the main tools of theoretical and numerical turbulence
research (Monin and Yaglom, 1975).  The same can be said about the
dimensionality of the problem; indeed, many natural flows that span
large number of scales possess features of both three- and 
two-dimensional turbulence and can be classified somewhere 
between the purely 3D and 2D extremes.  Thus, even though the focus
of this paper is isotropic 2D turbulence one should keep in mind that
applications of the results to quasi-2D flows are thought.
Quasi-2D turbulent flows are widely found in geophysics and engineering.
Although under normal circumstances all flows are unstable 
to three-dimensional instabilities (Batchelor, 1969), there exist natural 
situations when a flow may attain a quasi-2D configuration or even 
become quasi-two-dimensionalized on certain scales. 
There are two major factors that may cause a flow to become quasi-2D:
geometry of the flow boundaries and/or certain body forces (or `extra 
strains') whose action leads to smoothing of the velocity 
fluctuations in a preferred direction.  While in the geophysical 
context both factors are equally important (i.e., the small aspect
ratio, density stratification, rotation), in the engineering context
the second factor usually pre-dominates (for instance, the so called 
mechanism of `magnetic friction' in magnetohydrodynamic
flows with low magnetic Reynolds number; Sommeria and Moreau, 1982).

Although mathematical modeling of quasi-2D flows has important practical
applications, particularly in the atmospheric and oceanic sciences, it
has not received as much attention in the literature as the modeling of
the 3D flows.  Partly, it can be explained by the fact that the quasi-2D
problems are less computationally intense than their 3D counterparts.  Thus
there exists a hope that in the near future, practically important quasi-2D
problems can be solved using DNS in which all scales are resolved (Lesieur, 1990). 

In addition, despite the specific
peculiarities of quasi-2D flows related to the energy and vorticity dynamics, 
their subgrid scale representation has not received sufficient attention
so far.  There have been attempts to parameterize the SGS processes in
quasi-2D flows similarly to those in 3D flows using Laplacian or biharmonic 
dissipation, the most advanced method being the anticipated potential 
vorticity method (Sadourny and Basdevant, 1985).  However, such methods can only
perform well in the vorticity dissipation subrange when energy is injected
on relatively large scales.  Being applied in the energy transfer subrange, 
they will lead to energy dissipation, contradicting basic energy and 
vorticity transfer dynamics of quasi-2D turbulence.  Moreover, LES of quasi-2D 
flows in the energy transfer subrange has never been attempted, despite the 
fact that such flows would bear strong analogy to large scale oceanic and 
atmospheric circulation and that DNS of such flows cannot be expected in the
foreseeable future.  Thus, there exists a need to improve our understanding
of the SGS processes in the energy transfer subrange of quasi-2D flows and
to successfully simulate such flows when their energy sources reside
in the SGS region.  These both issues are addressed in the present paper.
In the next section, the basic difficulties of the SGS representation of 
the quasi-2D flows in the energy transfer subrange are discussed.  Then, the 
following section elaborates on the notion of the two-parametric viscosity
and explains how this viscosity resolves the conflict between inverse
transfer of energy and direct transfer of enstrophy.  In section 4, 
advantages and deficiencies of various implementations of the two-parametric 
viscosity for LES of 2D turbulence in the energy subrange are described.
In section 5, simplified SGS representations for LES of 2D turbulence are
considered and the notion of the stabilized negative viscosity (SNV) is
introduced.  Finally, section 6 discusses the results and provides some
conclusions.

\head {2. Basic problems of the SGS representation of quasi-2D flows in the
energy transfer subrange}

Confined to two dimensions, turbulent flows become non-vortex-stretching 
and undergo dramatic structural changes (Kraichnan and Montgomery, 1980). The 
most profound modifications take place in the dynamics of energy and vorticity 
transfer. It is well known that in isotropic homogeneous 3D turbulence, 
the direct energy cascade from large to small scales facilitates efficient 
energy dissipation by molecular viscosity. This process is accompanied by and 
closely related to the production  of enstrophy (mean square vorticity) through 
vortex stretching mechanism. Since in 2D flows vortex stretching can not occur, 
the enstrophy then is conserved.  Thus, 2D inviscid fluids possess two nontrivial 
integrals of motion: the energy and the enstrophy. The enstrophy conservation 
prevents cascade of energy from large to small scales because such cascade 
would increase enstrophy (Kraichnan, 1967, 1971; Kraichnan and Montgomery, 1980).

Mathematically, this important feature is illustrated by the Fj\o rtoft theorem 
(Lesieur, 1990).  However, the direct cascade of enstrophy from large to small 
scales is possible. It results in molecular dissipation of large scale vorticity at 
small scales.  Drawing analogy to eddy viscosity in 3D flows, one can infer that 
small scale processes in 2D flows generate an {\it effective}, or eddy viscosity 
for the {\it vorticity} of large scales.  However, introducing an eddy viscosity
concept in 2D flows seems to be intrinsically inconsistent and self-defeating 
because the dissipation of enstrophy will be accompanied by the dissipation 
of energy, which is physically incorrect.  This controversy calls for 
modification of the eddy viscosity concept for quasi-2D flows; the issue 
in the focus of the present paper. 

More detailed consideration of the transport processes in 2D turbulence
reveals that they depend on the wave numbers of the energy injection,
$k_f$.  For $k < k_f$, the energy cascades up scales (inverse cascade),
while the enstrophy flux is zero.  For $k > k_f$, the energy flux is
zero, but there exists the direct enstrophy flux (Kraichnan, 1971).  If
LES of a quasi-2D flow is thought, the proper SGS parameterization
should depend on the wave number of the energy source, $k_f$, i.e.,
whether $k_f$ belongs in the resolved (or explicit) or unresolved (or SGS)
region.  In the former case, a simple hyperviscous SGS representation
may suffice, because it should only account for the enstrophy dissipation
due to the direct cascade.  However, if the forcing is located in the
subgrid scales, then the hyperviscous SGS representation would lead to
erroneous results since it will constitute energy dissipation in
non-energy-dissipating flows.  To sustain such flows, one would need
to introduce a large scale energy source in the energy cascade subrange.
A possible solution to this problem would be to replace an SGS forcing 
by a forcing located in the explicit region near $k_c$, where $k_c$ is the
cutoff wave number corresponding to the grid resolution.
However, this solution is not only quite cumbersome but it also 
significantly distorts the explicit scales near $k_c$. In addition, 
this approach is difficult for implementation in the physical space, 
particularly for bounded systems and/or systems with spatially 
non uniform energy sources.

Another solution would be to introduce a negative eddy viscosity, first  
extensively discussed by Starr (1968), as an SGS parameterization of the 
unresolved energy source.  Recent studies of flows with negative viscosity 
were conducted by Gama et al. (1991,1994) in 2D and Yakhot and 
Sivashinsky (1985) in 3D. Although such an SGS 
representation could address the issue of the inverse energy cascade,
it would not satisfy the constraint of the zero enstrophy flux.
In addition, equations of motion with negative viscosity produce ill
posed problems.  It appears therefore that addressing the issue of SGS
representation for quasi-2D flows in self-consistent and comprehensive way 
would require full consideration of energy and enstrophy dynamics and
should be based upon the corresponding transport equations.  Such an
approach was first outlined by Kraichnan (1976) who introduced the notion
of two-parametric viscosity.  This approach and its implications will
be elaborated in the next section.
\head {3. Two-parametric viscosity as SGS representation of quasi-2D flows}

Two-dimensional incompressible turbulent flows are described by the 
vorticity equation  

$$
{\partial \zeta \over \partial t} + {\partial \left (\nabla^{-2} \zeta, 
\zeta \right ) \over \partial (x, y) }   
= \nu_0 \nabla^2 \zeta + f,  \eqno(1)
$$
where $\zeta$ is fluid vorticity, $\nu_0$ is molecular viscosity
and f is external force.

The introduction of the classical eddy viscosity concept for LES with
Eq. (1) implies that there is a distinct scale separation between the
resolvable and SGS modes.  Indeed, only if such a separation exists, the
eddy viscosity would be $k$-independent and a function of the cutoff wave
number $k_c$ only.  However, the assumption of scale separation fails in
all turbulent flows, particularly in 2D flows, such that, strictly speaking, 
an SGS representation should depend on two parameters, $k$ and $k_c$. 
Such two-parametric viscosity, denoted by $\nu(k\vert k_c)$, was first
introduced by Kraichnan (1976).   It describes the energy exchange 
between given resolved vorticity mode with the wave 
number $k$ and all SGS modes with $k > k_c$.  The two-parametric viscosity
is derived from the evolution equation for the spectral enstrophy density 
$\om(k,t)\equiv (4 \pi)^{-1} k \langle\zeta({\bf k},t)\zeta(-{\bf k},t)\rangle$, 
where $\langle\ldots\rangle$ denotes ensemble averaging: 

$$
\left(\partial_t + 2 \nu k^2\right) \om (k,t) = {\cal T}_{\om}(k,t). 
\eqno(2)
$$
The enstrophy transfer function in (2), ${\cal T}_{\om}(k,t)$, 
is given by

$$ 
{\cal T}_{\om}(k,t)= {k \over 2 \pi} \Re \left \{ \int_{{\bf p}+{\bf q}={\bf k}}
{{\bf p} \times {\bf q} \over {p^2}}
\langle\zeta({\bf p},t)\zeta({\bf q},t)\zeta(-{\bf k},t)\rangle
{d{\bf p}~d{\bf q} \over {(2\pi)^2}} \right \}. \eqno (3) 
$$

For a system in statistical steady state the two-parametric transfer 
${\cal T}_{\om}(k\vert k_c)$ and viscosity $\nu(k\vert k_c)$ are 
calculated from (3) by extending integration only over all such triangles 
$({\bf k, ~p, ~q})$ that $\vert k-p \vert < q < k+p$ and $p$ and/or $q$ 
are greater than $k_c$:

$$
\nu (k\vert k_c)= - {{\cal T}_{\om}(k\vert k_c)\over 2 k^2 \om(k)}. 
\eqno (4)
$$

In a wide class of quasi-normal approximations ${\cal T}_{\om}(k\vert k_c)$ 
in two dimensions is given by

$$ \eqalignno {
{\cal T}_{\om}(k \vert k_c) &= {2 k \over \pi} \int \!\int_\Delta 
\Theta_{-k,p,q} (p^2-q^2) \sin\alpha \left [{p^2-q^2 \over {p^3 q^3}} 
\om(p) \om(q) \right .\cr
&- \left . {k^2 - q^2 \over 
{k^3 q^3} } \om(q) \om(k) + {k^2 -
p^2 \over {k^3 p^3}} \om(p) \om(k) \right ]~dp dq, & (5)\cr  }
$$
where $\Theta_{-k,p,q}$ is the triad relaxation time. 
Also, the angle $\alpha$ is formed by the vectors ${\bf p}$ and ${\bf q}$, 
and $\int \!\int_\Delta$ denotes integration over the area defined above. 

Different spectral closure models provide different specification
of $\Theta_{-k,p,q}$.  Chekhlov et al. (1994) compared $\nu (k \vert k_c)$ 
calculated from their DNS data with 512$^2$ resolution with those evaluated 
by Kraichnan (1976) using his Test Field Model (TFM) and obtained from
the Renormalization Group (RG) theory (see Appendix).  The results
of this comparison are shown in Figs. 1 and 2.  Figure 1 presents the
DNS-inferred normalized two-parametric viscosity, 
$$
N(k/k_c) \equiv \nu(k \vert k_c) / \vert \nu(0 \vert k_c) \vert,  \eqno(6)
$$
along with the TFM- and RG-based analytical predictions.  The results are in 
a very good agreement with each other over 
the entire energy transfer range, up to the wave numbers close to $k_c$, 
where the DNS data saturates, while TFM and RG curves exhibit sharp cusp.  
This theoretical cusp is due to the fact that as $k \to k_c$, more and more
elongated triads with either $p$ or $q \ll k_c$ become involved in the 
energy exchange between the mode $k$ and the subgrid scale modes.
The contribution from these triads to the energy exchange near $k_c$ is
very significant and results in the cusp behavior. However, in finite box 
DNS with large-scale energy removal, the energy of small wave number modes 
is reduced and $\nu(k \vert k_c)$ is expected to saturate near $k_c$.  
Indeed, when the RG-based $\nu(k \vert k_c)$ was re-calculated based upon
the DNS energy spectrum, the unnormalized DNS- and RG-based two-parametric 
viscosities were found to be in very good agreement for all
wave numbers, as shown in Fig. 2.  

Figures 1 and 2 show that for the large scale modes, for which $k \ll k_c$
and scale separation exists, the effect of the SGS modes is represented by a
negative and constant viscosity, such that these modes {\it gain}
energy from their SGS counterparts by means of the inverse transfer.
On the other hand, $\nu(k \vert k_c) > 0$ for $k \to k_c$ such that the 
modes close to $k_c$ lose their energy to the SGS modes.  The difference
between the large scale gain and the small scale loss is equal to $\eps$,
the rate of the energy input due to the forcing $f$ in Eq. (1); see also
Eq. (13) below.  If enstrophy balance is considered, recall that the 
enstrophy transfer is most efficient at small scales, such that the 
resulting balance for the resolvable scales turns out to be zero [Kraichnan, 
1971; see also Eq. (18) below], i.e., the enstrophy is conserved.  This 
explains how the two-parametric viscosity resolves the controversy of the 
inverse cascade of energy and conservation of enstrophy in the energy 
subrange of 2D flows.  It appears therefore that the only physically 
correct way to represent SGS processes in 2D turbulence would be through 
the two-parametric viscosity.  Such an approach has become quite popular 
in simulations of 3D flows (Domaradzki et al., 1987; Galperin and Orszag, 1993)
but has not yet been fully explored for LES of quasi-2D flows.  The implementation
of the two-parametric viscosity for LES of 2D turbulent flows, the arising 
problems, their solutions and results are described in the following sections.  

\head {4.  Implementation of the two-parametric viscosity for LES of 2D
turbulence}

To test the two-parametric viscosity-based SGS parameterization in the energy 
transfer subrange, a series of LES of 2D turbulence in Fourier space was 
designed.  These LES were based upon Eq. (1) in which all subgrid scale 
processes including the forcing were represented by the two-parametric 
viscosity $\nu(k \vert k_c)$,        
$$
{\partial \zeta ({\bf k}) \over \partial t} + 
\int_{|p|,|k-p| < {k_c}}
{{\bf p} \times {\bf k} \over {p^2}}
\zeta({\bf p})\zeta({\bf k} - {\bf p})
{d{\bf p}\over {(2\pi)^2}}
=  \nu(k \vert k_c) k^2 \zeta({\bf k}), ~~~~~ 0 < k < k_c.  \eqno(7)
$$
It is important to reiterate that in LES of 2D turbulence in the energy
subrange, the source of energy resides on the unresolved scales, such
that Eq. (7) appears unforced.  However, as was explained earlier,
the negative part of $\nu(k \vert k_c)$ serves as the only energy source for 
the resolved modes.  In the course of the present LES it was 
found that numerical results critically depend on the way $\nu(k \vert k_c)$
is introduced in the solver.  Thus, a series of simulations was designed with
the purposes of understanding the nature of the problems associated with the
implementation of the two-parametric viscosity and of identifying the most 
viable and robust ways to use this viscosity in LES of 2D flows.

\subhead{4.1.  Description of numerical method}   

The numerical solver used in the present calculations was based upon 
Fourier-Galerkin pseudo-spectral formulation (Orszag, 1969) in the periodic 
box $[0,2\pi] \times [0,2\pi]$ and was essentially the same as that utilized 
in the preceding DNS (Chekhlov et al., 1994). The present LES employed 
162$^2$ resolution with complete dealiasing based on the 2/3-rule; the cutoff 
wave number was set at $k_c$ = 50, which is about half of the resolution used
in DNS of Chekhlov et al. (1994).  The preprocessing was done using the 
second order Runge--Kutta scheme, while marching in time was accomplished 
using an implicit, second order, stiffly 
stable Adams scheme (Karniadakis et al., 1991).  The initial flow field was set 
to zero everywhere except for a narrow band of wave numbers in the middle part of
the spectrum where it was assigned random Gaussian distribution.  The DNS
inferred
value of $\overline{\epsilon}$ was about $5.19\times 10^{-10}$; the same 
$\overline{\epsilon}$ was used in LES.  The time step in LES was set to 
$\delta t=0.5$ satisfying both convective and viscous necessary 
conditions for linear stability.   Based upon the size of the largest energy
containing eddy with the wave number $k_{\rm {min}}$, $2 \pi/k_{\rm {min}}$, 
and the total energy of the steady-state ${\overline  E}(t)$, the 
maximum large scale eddy turnover time defined as $\tau_{tu} = 2 \pi /
(k_{\rm {min}} \sqrt{2{\overline  E}})$ was about $\tau_{tu} \simeq 3600$ for
LES of cases 1 and 2 below, where $k_{\rm {min}} = 1$, and $\tau_{tu} \simeq 900$
for cases 3,4, and 5, for which $k_{\rm {min}} = 4$.

\subhead{4.2. Case 1.  Flow independent $\nu(k \vert k_c)$}

With $\nu(k \vert k_c)$ known and flow independent, Eq. (7) can be solved directly.
According to Fig. 1, $\nu(k \vert k_c)$ can be obtained from DNS or from some 
statistical theory of turbulence.  Thus, in the first LES numerical solver 
for Eq. (7) utilized $\nu(k \vert k_c)$ derived from the renormalization 
group (RG) theory of turbulence (see Appendix for the details).
As shown in the Appendix,
$$
\nu(k \vert k_c) =  0.327 \eps^{1/3} k_c^{-4/3} N(k/k_c),   \eqno(8)
$$
where $N(k/k_c)$ is given by (6), and since $\eps$ is a constant, 
$\nu(k \vert k_c)$ is a function of $k$ and $k_c$ only.  It was assumed 
that thus defined $\nu(k \vert k_c)$ 
would be capable of supporting inverse energy cascade with constant SGS 
energy input $\eps$.  To verify this assumption, one needs to examine the 
evolution of total energy and enstrophy of the resolved modes, 
${\overline E}(t)$ and ${\overline \om}(t)$, respectively.  By definition, 
${\overline E}(t) = \int_0^{k_c} E(k,t) dk$, where 
$E(k,t) \equiv (4 \pi k)^{-1}\langle\zeta({\bf k},t)\zeta(-{\bf k},t)\rangle$
is the spectral energy density, and ${\overline \om}(t) = \int_0^{k_c} 
\om(k,t) dk$.  The basic requirement to LES would be that ${\overline E}(t)$ 
and ${\overline \om}(t)$ of LES have the same behavior as those derived from
Eq. (1) for which the evolution laws are ${\overline  E}(t) \propto \eps t$
and ${\overline \om}(t) =$ const, due to the conservation of the inviscid 
integrals for ${\overline E}(t)$ and ${\overline \om}(t)$ (recall that in the
energy subrange of 2D turbulence the rate of the enstrophy flux $\eta = 0$).  
Figures 3a,b show that in the first LES, both ${\overline E}(t)$ 
and ${\overline \om}(t)$ exhibit nonlinear growth indicating that not only 
the rate of the energy transfer to the resolvable scales $\epsl \ne$ const, but also 
the rate of the enstrophy transfer $\etal \ne$ 0.  Figure 4 shows that instantaneous 
spectrum $E(k,t)$ also reveals tendency to growing up with time without stabilizing 
around any universal distribution.  At some point of its evolution, $E(k,t)$ crossed
the 
Kolmogorov law
$$
E(k) = C_K \eps^{2/3} k^{-5/3},  \eqno(9)
$$
where $C_K \approx 6$ is the Kolmogorov constant, and $\eps$ was close
to its prescribed DNS value.  However, at larger times the Kolmogorov 
scaling (9) was lost, while $\epsl$ kept growing. The roots of 
the problem are revealed when one calculates the rate of the energy input 
into all resolved modes, $\epsl$. The energy equation derived from the 
definition of spectral energy density and Eq. (7) yields
$$
\epsl(t) \equiv {\partial {\overline E}(t)\over \partial t} = 
- 2 \int_0^{k_c} \nu(k \vert k_c) E(k,t) k^2 dk.  \eqno(10)
$$
Since in the first series of LES, $\nu(k \vert k_c)$ depends on $k$ and 
$k_c$ only, and $E(k,t)$ is a dynamic variable that depends on the 
evolution of the flow field, $\epsl(t)$ also turns out to be time dependent. 
This is not only in direct conflict with the requirement that $\epsl = \eps =
$const, but also leads to a positive feedback between the energy input 
and total energy of the system, which results in numerical instability. 
To correct this problem, one must ensure that $\epsl =$ const.  This can be 
achieved by allowing $\nu(k \vert k_c)$ to become time dependent and 
related to resolved variables.  The philosophy of using actual flow 
field characteristics to determine energy input and dissipation would
be analogous to a standard practice of 3D LES. 

Inspection of Eq. (10) reveals that $\epsl(t)$ is proportional to the total 
enstrophy of the resolvable field, such that stipulating $\epsl =$ const would
require $\nu(k \vert k_c)$ to become time dependent and inversely proportional 
to $\omb(t)$ (the time dependency of $\nu(k \vert k_c)$ will be implied in the 
following discussion but suppressed in notations).  To find the explicit
dependency of $\nu(k \vert k_c)$ on $\omb(t)$, let us assume that 
$\nu (k \vert k_c)$ can be represented by 
$$
\nu (k \vert k_c) = F(\omb) N(k/k_c),    \eqno(11)
$$
where $F(\cdot)$ is some function of $\omb(t)$ and $N(k/k_c)$ is still 
given by (6). According to  Fig. 1, $N(k/k_c)$ can be split into 
negative and positive terms,
$$
N(k/k_c) = -1 + \phi(k/k_c),~~ \phi(k/k_c) > 0,~~0 \le k/~k_c \le 1.  
\eqno(12)
$$
One can now calculate integral in (10) using Eqs. (11) and (12): 
$$
\eqalignno {
\epsl &= 2 F(\omb) \int_0^{k_c} E(k,t) k^2 dk -
2 F(\omb) \int_0^{k_c} \phi(k/k_c) E(k,t) k^2 dk \cr
&= 2 F(\omb) \omb(t) - 2 F(\omb) \int_0^{k_c} \phi(k/k_c) E(k,t) k^2 dk.  
&(13) \cr  }
$$
Integration of (13) for the Kolmogorov spectrum with $\phi(k/k_c)$ 
evaluated from the RG theory (see Appendix) yields
$$
\epsl \simeq 0.8 F(\omb) \omb(t).  \eqno(14)
$$
Equation (14) shows that to satisfy the requirement $\epsl = \eps =$ const,
one has to impose
$$
F(\omb) = \eps/ [0.8 \omb(t)],  \eqno(15)    
$$
such that the two-parametric viscosity (11) becomes
$$
\nu (k \vert k_c) = {\eps \over 0.8 \omb(t)} N(k/k_c) = - {\eps \over 0.8 
\omb(t)} + {\eps \over 0.8 \omb(t)}  \phi(k/k_c).   \eqno(16)
$$
The first term in the right hand side of (16) accounts for the SGS energy 
input while the second term represents the high wave number dissipation as
$k \to k_c$.  As was argued earlier, to ensure $\epsl =$ const, the energy 
source term must be time dependent and inversely proportional 
to $\omb(t)$.  Thus, the  negative feedback between energy input and enstrophy 
of the resolved modes is the mechanism that stabilizes numerical process. 
Note that the SGS formulation based upon Eq. (16) complicates
Eq. (7) because its right hand side now depends on the functional of the 
solution, $\omb(t)$. However, on the one hand, it is clear from the 
presented analysis that LES of 2D turbulence based upon Eq. (7) is impossible 
if $\nu (k \vert k_c)$ depends on $k$ and $k_c$ only.  On the other, the SGS 
representation (16), though complicated, is in line with the eddy 
viscosity approach, in which the eddy viscosity coefficient is usually 
solution dependent (Smagorinsky, 1963,1993; Yakhot and Orszag, 1986; Yakhot et al.,
1989).

\subhead{4.3. Case 2.  Flow dependent $\nu(k \vert k_c)$ with no large scale drag}

The SGS formulation (16) was used in the second LES and considerable 
improvement over Case 1 was observed.  Figures 5a,b show that up to 
the simulated time $t = 3 \tau_{tu}$, ${\overline  E}(t)$ grows 
linearly, while $\omb(t)$ attains a constant value.  Figure 6 shows that 
during the same $t$, $E(k,t)$ quickly approaches steady state Kolmogorov 
distribution (9).  However, for $t > 3 \tau_{tu}$ the flow field 
undergoes irreversible modifications; the behavior of ${\overline  E}(t)$ and 
${\overline  \om}(t)$ changes, while $E(k,t)$ begins to deviate from the 
Kolmogorov law.  All these changes reflect the basic problem of the present
LES that simulates the behavior of an infinite system in a finite computational 
box (Smith and Yakhot, 1993a,b). In this box, the smallest wave number modes 
become energy saturated at $t \simeq 3 \tau_{tu}$, and, if the energy of these 
modes is not removed, they begin to alter the behavior of the entire flow field.
Therefore, to extend LES beyond $t \simeq 3 \tau_{tu}$, one needs to prevent the
accumulation of energy at the lowest modes, which was accomplished in LES 
of Case 3.  Note however that by the time $t \simeq 3 \tau_{tu}$ the inverse
cascade swept through all the resolved modes such that they became energy
saturated and attained the steady state.  Therefore, one should expect that
in LES with $t > 3 \tau_{tu}$ both ${\overline  E}(t)$ and ${\overline  
\om}(t)$ remain nearly constant.

\subhead{4.4. Case 3.  Flow dependent $\nu(k \vert k_c)$ with large scale drag}

The simplest way to withdraw energy from the lowest modes would be by mere
setting 
to zero the amplitudes of those modes.  However, such a ``chopping'' alone is 
known to produce unsatisfactory results (Browning and Kreiss, 1989).  
Therefore, in addition to the chopping, one needs to introduce a mechanism 
that would account for the energy exchange between the
resolved modes and the low wave number modes excluded in LES.  Such a
mechanism,
the large scale drag, was introduced in this study in analogy to the two-parametric
viscosity.  However, the derivation of this large scale drag is beyond the scope of
this paper and will be presented elsewhere.  

The large scale drag was implemented in the third LES, whereas the amplitudes
of all modes with $k < k_{\rm {min}} = 4$ were set to zero.  As was explained in
section 3 and shown in Fig. 2, such a chopping results in flattening of the
cusp in $N(k/k_c)$ as $k \to k_c$, such that this function had to be 
recalculated which in turn led to modification of the coefficient in (14)
from 0.8 to 0.87.  The results of the third LES are shown in Figs. 7a,b and 
8a,b.  One can see that this simulation could virtually be extended 
indefinitely, with $\Eb(t)$ and $\omb(t)$ slightly oscillating around 
their steady state values (the source of these oscillations is probably 
the self-adjustment of the numerical scheme to the mismatch between the small scale
energy forcing and large scale withdrawal).  The instantaneous energy
spectrum, Fig. 8a, exhibits steady and nearly perfect Kolmogorov scaling.
Since the large scale drag enables one to dramatically increase the integration
time in LES, it will be retained in all further simulations.  Note however
that these simulations will pertain to steady state rather than time
developing flows.  

An important characteristic of 2D turbulence in the energy transfer subrange is 
the energy flux $\Pi_E(k) = \int_0^k {\cal T}_\Omega(n) n^{-2} dn$ 
which theoretically should be equal to $\eps$ for any $k$.  Note that the enstrophy
transfer function ${\cal T}_\Omega(n)$ is defined by (3) where integration
is extended over all triangles $\k + \p + \q = 0$ including the SGS ones
for which $k$ and/or $p$ and/or $q > k_c$.  However, if $\Pi_E(k)$ is 
calculated for LES results, then the SGS triangles should be excluded such 
that the integration area in (3) is reduced.  On the other hand, the energy 
flux into the interval $0 < n < k$ in LES consists of two contributions, 
$\int_0^k {\cal T}^\prime_\Omega(n) n^{-2} dn$ (where ${\cal T}^\prime_\Omega$ 
implies the reduced integration area in (3)), and the direct flux from the 
SGS modes, $-2 \int_0^k \nu(n \vert k_c) n^2 E(n) dn$.  Figure 8b shows that 
thus calculated $\Pi_E(k)$ remains nearly constant for $k > 16$ and is equal 
to about $5 \times 10^{-10}$, which is very close to the specified value of 
$\eps$ obtained from DNS by Chekhlov et al. (1994). The decrease of 
$\Pi_E(k)$ at small $k$ is due to the large scale energy removal.

\subhead{4.5. Case 4.  Flow dependent energy input with flow independent
dissipation}

It would be tempting to simplify (16) by relaxing the time dependency in the
dissipation term.  It is not clear a priori whether or not this time dependency is
critical, and to find out about it a fourth LES was conceived in which SGS 
representation (16) was modified by replacing $\eps / 0.8~\omb(t)$ in the 
dissipation term by the RG derived expression $0.327 \eps^{1/3} k_c^{-4/3}$.  
Figures 9a,b and 10a,b,c show that this simplified SGS scheme performs in a very 
robust way with no oscillations at all for a relatively long time, 
$t \simeq 30 \tau_{tu}$.  Similarly to Case 3, there exists
a mismatch between the small scale forcing and large scale energy removal, but
since the dissipation in Case 4 cannot self-adjust, the solution begins to deteriorate
when this mismatch accumulates significantly.
Still, Case 4 LES could be extended to many more turnover times than the 
corresponding DNS, the result quite remarkable for its own sake.  During this
time, the instantaneous energy spectrum, Fig. 10, exhibits steady and nearly
perfect Kolmogorov scaling almost indistinguishable from that in Fig. 8.
Figure 10b shows that the Kolmogorov constant $C_K$ in (9) is about 5, in good
agreement with the RG derived value of 5.12 (see Appendix).  As in Case 3,
the energy flux $\Pi_E(k)$ shown in Fig. 10c is almost constant for $k > 16$
and is about $-5 \times 10^{-10}$.

Although the SGS formulation (16) is relatively easy to implement in
spectral LES,  in practical applications, particularly in the physical space, 
it would be far more useful to approximate $N(k/k_c)$ analytically. 
An additional benefit of such an approach would be a possibility to carry out
further analytical studies of this SGS representation; the direction pursued
in the next section.  

\head {5.  Stabilized negative viscosity (SNV) formulation}

For practical implementation of SGS formulation (16) it is convenient to 
approximate $N(k/k_c)$ in (6) by a series in powers of $k^2$. 
It was found that even the first two terms of this series, 
$$
N(k/k_c) \simeq  -1 + \alpha (k/k_c)^2,  \eqno(17)
$$
where $\alpha$ is a constant, are sufficient to perform successful LES of 
2D turbulence.  To find $\alpha$ recall that representation (17) must 
ensure zero enstrophy transfer in the energy subrange,
$$
\etal = 2 \int_0^{k_c} \nu(k \vert k_c) E(k,t) k^4 dk = 
2 F(\omb) \int_0^{k_c} N(k/k_c) E(k,t) k^4 dk = 0. \eqno(18) 
$$
Substituting (17) into (18) and assuming that $E(k,t)$ is Kolmogorovian, 
one finds that $\alpha = {8 \over 5}$ such that (13) becomes
$$
\eqalignno{ 
\epsl &= - 2 F(\omb) \int_0^{k_c} [-1 + {8 \over 5}(k/k_c)^2] E(k,t) k^2 dk \cr
&= 2 F(\omb) \omb(t) - {16 \over 5} F(\omb) k_c^{-2} \Pb(t),  &(19) \cr}
$$
where $\Pb(t) \equiv \int_0^{k_c} E(k,t) k^4 dk$ is the total palinstrophy of the
resolved modes.  For Kolmogorovian $E(k,t)$, (19) can be integrated analytically
to yield $\Pb(t) = {2 \over 5} \omb(t) k_c ^2$ and
$$
F(\omb) = {25 \over 18}~ {\eps \over \omb(t)}.   \eqno(20)
$$
Equation (20) completes the two-term SGS parameterization for LES of 2D 
turbulence in which the right hand side of Eq. (7) takes the form
$$
\nu (k \vert k_c) k^2 \zeta ({\bf k}) = {25 \over 18}~ {\eps \over \omb(t)}
\left [ -1 + {8 \over 5} \left ( {k \over k_c} \right)^2 \right ] 
k^2 \zeta ({\bf k}). \eqno(21)
$$
As in (16), the first term in the right hand side of (21) accounts for the 
energy flux from the unresolved modes and the inverse cascade of this energy, 
while the second term represents the energy dissipation near the cutoff.  
Using (21) and following the philosophy of Cases 3 and 4 LES, two more
simulations were designed, with flow dependent and independent dissipation
term in (21).

\subhead {5.1. Case 5.  Two-term LES with flow dependent dissipation}

Simulations performed with the formulation (21) slightly adjusted to account for
the finiteness of the computational domain are shown in Figs. 11a,b and 12.  They
exhibit very little difference compared to the third LES that employed the full 
curve $N(k/k_c)$ given by (6) and shown in Figs. 7a,b and 8.  One infers therefore 
that (21) is a viable two term SGS representation for LES of 2D turbulence; 
obviously, (21) is significantly simpler than (16).  

\subhead {5.2. Case 6.  Two-term LES with flow independent dissipation}

For practical purposes, it would be most appealing to use formulation (21) with
the dissipation term constant.  A numerical experiment analogous to that of Case 
4 LES was conducted with the energy source in (21) not changed but in the 
dissipation term, $\omb(t)$ was replaced by its value calculated for the Kolmogorov 
spectrum (9).  Such an approach yields the dissipation term in (21) in the form
$$
A k^4 \zeta ({\bf k}),   \eqno(22)
$$
where $A$ is a constant given by
$$
A = 0.511 \eps^{1/3} k_c^{-10/3},  \eqno(23)   
$$
which corresponds to $C_K$=5.8.  The results of this  case 5 LES 
are presented in Figs. 13a,b and 14a,b; there is a very good agreement with the 
corresponding results obtained with the full curve $N(k/k_c)$ up to $t \simeq 40 
\tau_{tu}$ after which, similarly to Case 4, the solution begins to deteriorate.  
The Kolmogorov constant $C_K$ shown in Figs. 14b is close to 5.8, which is
consistent with derivation of (23).
One can therefore infer that (21), (23) is probably the simplest SGS representation 
for LES of 2D turbulence possible.

Further advantages of the SGS representation (23) are revealed when LES of 
quasi-2D turbulence is sought in the physical space, where Eq. (21) combined 
with (23) leads to the following SGS representation:
$$
- {\partial \over \partial x_i} \left ( A_2  {\partial \over \partial x_i} 
\zeta({\bf x})\right ) - A_4 {\partial^4 \over \partial x_i^2 \partial x_j^2 } 
\zeta({\bf x}),    \eqno(24)
$$
where
$$
\eqalignno {
A_2 &= {25 \over 18}~ {\eps \over \omb({\bf x})},  &(25) \cr
A_4 &= A = 0.511 \eps^{1/3} (\Delta /2 \pi)^{10/3} = {\rm const}, &(26)  \cr }
$$
and where $\omb({\bf x})$ denotes the enstrophy averaged over an area adjacent to the 
grid cell, and $\Delta$ is the grid resolution (note that the Laplacian term in (24) 
is written in the conservative form).  Equation (24) thus includes two terms, the
{\it negative} Laplacian and positive (in the sense of dissipation) biharmonic, 
and structurally resembles the Kuramoto--Sivashinsky equation widely known from 
combustion theory (Sivashinsky, 1977) and flows with chemical reactions (Kuramoto 
and Tsuzuki, 1976; Kuramoto, 1984).  However, the SGS representation (24)-(26)
combined with the explicit equation for the resolved scales produces far more 
complicated equation than the Kuramoto--Sivashinsky equation because generally its 
coefficients are not constant but, as in the eddy viscosity approach, are functions of 
the flow.  There have been previous attempts to use formulation similar to 
(24)-(26) but with constant coefficients 
(dubbed the Kuramoto--Sivashinsky--Navier--Stokes equation, 
see Gama et al., 1991) to perform LES of 2D turbulence.  However, they were not 
overly successful even in reproducing  the Kolmogorov spectrum, mostly because they 
used constant ``eddy viscosity'' coefficients.  Since, on the one hand, SGS 
representation (24)-(26) includes a negative Laplacian viscosity term and a positive, 
stabilizing, dissipation term, but, on the other hand, it is quite different from 
the Kuramoto--Sivashinsky equation in which ``viscosity'' coefficients are constant, 
it will be referred to as the Stabilized Negative Viscosity (SNV) formulation.  
The SNV formulation is expected to be particularly useful in simulations of
atmospheric and oceanic flows where large scale motions are quasi-two-dimensional 
by their nature while small scale forcing is significant (Monin, 1986).

\head {6.  Discussion and Conclusions}

LES of homogeneous and isotropic 2D turbulence in the energy transfer subrange 
and appropriate SGS representation have been the central subject of this paper.  
Conservation of energy and enstrophy in 2D turbulent flows leads to coexistence
of two spectral transfer processes, upscale for energy and downscale for enstrophy.
Conventional eddy viscosity formulations are purely dissipative and fail to
accommodate both transfers simultaneously.  It is argued that proper SGS
parameterization
for LES of 2D flows is given by the two-parametric viscosity $\nu (k \vert k_c)$
introduced by Kraichnan (1976) that accounts for the energy (or enstrophy) exchange 
between given resolved and all SGS modes and which includes negative and positive 
branches;  the negative branch of $\nu (k \vert k_c)$ represents the unresolved, 
small scale forcing and inverse cascade of energy, while the positive one represents 
the dissipation.  It was shown that the negative and positive parts of the
two-parametric viscosity play vital role in ensuring that all conservation 
and spectral transfer laws of 2D turbulence are satisfied.  Then, 
$\nu (k \vert k_c)$ was used in LES of 2D turbulence in the energy transfer 
subrange where the negative part of $\nu (k \vert k_c)$ 
was the only energy source.  The sensitivity of numerical results to 
the way of implementation of $\nu (k \vert k_c)$ in numerical schemes was studied.
It was found that if $\nu (k \vert k_c)$ is specified as a flow independent 
parameter then a positive feedback is established between the forcing and 
the total energy of the system leading to numerical instability.  Thus, 
another scheme was designed in which $\nu (k \vert k_c)$ was flow dependent 
but the rate of the energy input was kept constant.  This LES exhibited a 
very stable behavior consistent with analytical theories and DNS; Kolmogorov 
scaling was evident and robust, and all conservation and spectral transfer 
laws were fulfilled.  Then, a simplified SGS representation was advanced, 
in which only two terms were retained: one negative and proportional to $k^2$,
and one positive and proportional to $k^4$.  The viscosity coefficient
in the negative term served as the energy source and had to be flow
dependent, to ensure that the energy input remains constant.  However, since the 
$k^4$ term is mostly active at high wave number region near the cutoff wave
number $k_c$, and its main role is energy and enstrophy dissipation at 
$k \to k_c$, this term did not need to be flow dependent and could 
essentially be kept constant and evaluated from analytical theories of 
2D turbulence.  Indeed, LES with constant and flow dependent $k^4$ terms 
gave very similar results.  

The application of the two term parameterization to simulations in physical 
space results in the so called Stabilized Negative Viscosity (SNV) 
representation which includes negative Laplacian and positive biharmonic
viscosities.  Numerical implementation of the scheme with the constant 
biharmonic viscosity is obviously much simpler than that with the flow 
dependent viscosity.

On the one hand, the negative viscosity term is essential in SNV scheme;
on the other, this scheme substantially differs from the Kuramoto-Sivashinsky
equation with its constant viscosity because in SNV, the negative viscosity 
not only is not constant but is a nonlinear functional of the solution.
In fact, the SNV representation is a peculiar case of the eddy viscosity 
approach; one normally expects that if all scales of a turbulent flow are 
resolved then eddy viscosity would become equal to the molecular viscosity 
which is true in 3D turbulence.
In 2D turbulence, however, resolution of all scales must be accompanied by
restoration of the explicit small scale forcing which would result in
disappearance of both negative Laplacian and positive biharmonic viscosities
and appearance of one positive, Laplacian, molecular viscosity.

It is believed that the SNV representation should be especially useful 
in quasi-2D flows in which considerable amount of energy resides on the
unresolved scales.  Such flows are typical in geophysical fluid dynamics 
where, due to the topographic and other constraints, flows are quasi-2D and  
where the small scale forcing is the predominant source of energy.

\head {7.  Acknowledgments}

This research has been partially supported by ONR Grants N00014-92-J-1363, 
N00014-92-C-0118, and N00014-92-C-0089, NSF Grant OCE 9010851.  Stimulating
discussions with Dr. Ilya Staroselsky are appreciated. 

\head {8. APPENDIX:~~ DERIVATION OF FLOW PARAMETERS BASED ON RG
THEORY}

Fourier transformed Eq. (1) reads
$$ 
\zeta({\hat k}) = G^o({\hat k}) \int {{\bf k} \times {\bf q} \over q^2}
{\zeta({\hat q}) \zeta({\hat k} - {\hat q}) \over (2 \pi)^{3}}
d {\hat q} + G^o({\hat k}) f, \eqno ({\rm A}1)
$$
where ${\hat k} \equiv ({\bf k}, \omega)$, ${\hat q} \equiv ({\bf q},
\Omega)$, and $G^o({\hat k}) \equiv (i \omega + \nu_0 k^2 )^{-1}$ is the bare
propagator.  The random, zero-mean, white in time Gaussian stirring force 
$f$ in the right hand side of (A1) accounts for the steady energy input 
localized in the vicinity of a given wave number $k_f$. In the RG formalism 
(Yakhot and Orszag, 1986; Staroselsky and Sukoriansky, 1993)
the effect of a localized random stirring on the bare equation of motion
is equivalent to the effect of a spatially-distributed forcing with
the correlation function
$$
<f({\bf k},\omega)f({\bf k}^\prime,\omega^\prime)>
=D(k)(2 \pi)^{3} \delta({\bf k}+{\bf k}^\prime)
\delta(\omega+\omega^\prime) \eqno ({\rm A}2)
$$
on the renormalized equation.
Here, $D(k)=2 D_0 k^{-y+2}$, where $y$ and $D_0$ are different for 
$k \ll k_f$ and $k \gg k_f$. The region $k \ll k_f$ corresponds to the 
inverse cascade of energy for which $y=2$ and $D_0 \propto \overline \epsilon$, 
$\overline \epsilon$ being the constant energy injection rate; 
the region $k \gg k_f$ corresponds to the direct cascade of enstrophy
with the constant rate $\eta$; in that region $D_0\propto \eta$ and $y=4$.  
Note that in the framework of the 2D RG theory, the regions $k \ll k_f$ and 
$k \gg k_f$ are treated as separate asymptotic regimes.  Equation (A1) is 
defined in the interval $0 < k < \L$. 

Fourier coefficients of the vorticity field can be separated into
small scale, or fast modes, $\zeta^>(k)$, for which $\L-\D \L < k < \L$, 
and large scale, or slow modes, $\zeta^<(k)$, for which $0 < k < \L-\D \L$,
respectively.  The RG procedure detailed in Forster et al. (1977) and Yakhot and
Orszag (1986) consists of successive elimination of infinitesimal shells of the 
small scale modes from the equation of motion for the large scale modes. 
Essentially, $\zeta$ field decomposed into $\zeta^>(k)$ and $\zeta^<(k)$ 
parts is substituted into Eq. (A1); then, the one-loop approximation is invoked
and the average is taken over the small scale modes. Equation (A1) is then
rewritten in terms of $\zeta^<(k)$ and is defined in the abridged interval 
$0 < k < \L -\D \L$; it now contains a new term represented by the correction
to the Green function, 
$$
\delta G(\hk)^{-1}  = \int^>  { {k^2 q^2 - ( \k \cdot \q )^2 }
\over k^4 } \left ( \qsi - \qksi \right )~ G( \hk - \hq)~\vert G(\hq)
\vert ^2~D_o q^{-y+2} {d \hq \over (2 \pi)^{3}},   \eqno ({\rm A}3) 
$$
that accounts for the effect of the unresolved scales.  Here, $\int^>$ 
denotes an integration over the band of wave numbers being removed.  

The integral in (A3) is calculated in the limit $k \to 0, 
\, \w \to 0$.   The $O(k^2)$ terms provide correction to $\nu_0$. 
Iterating the scale elimination procedure, one can remove a finite band of
modes and obtain a differential equation for the parameter of renormalization
$\nu(k)$,  
$$
{ d \nu \over d k } = - {D_o \over 16 \pi \nu^2 k^{\e+1} },  \eqno  ({\rm A}4)
$$
where $\e = 4+y-d$.  For the energy inertial subrange considered
below, $y=2$ and $\e=4$.

The solution to Eq. (A4) is
$$
\nu(k) = \nu_0 \left \{ 1 + {3 D_o \over 64 \pi \nu_0^3 \L ^4}
\left [ \left ( {k \over \L}\right )^{-4} - 1 \right ]
\right \}^{1/3}.   \eqno ({\rm A}5)
$$
Assuming that $k/\L \ll 1$ and the molecular viscosity 
can be neglected $[\nu_0 \ll \nu(k)]$, one finds:
$$
\nu(k) = \left ( {3 D_o \over 64 \pi} \right )^{1/3} k^{-4/3}.   \eqno ({\rm A}6)
$$
It is argued (Forster et al., 1977;  Yakhot and Orszag, 1986) that iterated 
indefinitely, the RG procedure of small scale elimination converges to a fixed point 
solution whereas $\zeta(k)$ is described by the Langevin equation 
$$
\zeta(k, \omega) = G (k, \omega) f(k, \omega),  \eqno ({\rm A}7)
$$
where $G \equiv [i \omega + \nu (k) k^2]^{-1} $ is the renormalized
propagator.  The existence of the fixed point for RG procedure in 2D
has been proven by Staroselsky et al. (1995) for $\e \ll 1$.  Although the
feasibility of continuation of these results into the region of large
$\e$ is still an open question, it is assumed here that the fixed point 
solution (A7) exists at $\e =4$.

Equation (A7) allows one to calculate the vorticity correlator, 
$U(\k,\w) \equiv \langle\zeta({\bf k},t)\zeta(-{\bf k},t)\rangle$, 
as well as the kinetic energy spectrum, 
$$
E(k) = {1\over 4\pi}k^{-1}\int\,{d\w \over (2\pi)}\,U(k,\omega) = 
(3 \pi^2)^{-1/3} D_0^{2/3} k^{-5/3}. \eqno ({\rm A}8)
$$

Although $\nu (k)$ describes renormalization of the bare viscosity
$\nu_0$, it is not what is often comprehended as an eddy viscosity.
This is merely a response function of the nonlinear dynamical system
described by the renormalized Eq. (A7). As such, it allows for calculating
vorticity correlator and energy spectrum but does not directly relate to 
enstrophy and energy transfer and dissipation.  Furthermore, 
$[\nu (k) k^2]^{-1}$ can be viewed as a characteristic time scale of 
information loss at given ${\bf k}$ caused by nonlinear scrambling of 
all other modes (Dannevik et al., 1987).  Therefore, $\nu (k)$ has a 
meaning of an eddy damping parameter and is substantially one-point 
turbulence characteristic.

To analyze energy and enstrophy transfer, one needs to consider two-point
characteristics that account for interaction between a given explicit
mode $k < k_c$ and all modes $k > k_c$; $k_c$ is identified with the moving
dissipation cutoff. For this purpose, following Dannevik et al. (1987) the 
energy evolution equation should be derived at lowest nontrivial order of 
nonlinear coupling using fully renormalized propagator $G(k)$.
The resulting dynamical closure is similar to the Eddy-Damped, Quasi-Normal,
Markovian (EDQNM) approximation where the lowest order RG analysis is invoked
to obtain the eddy damping function.  The energy equation then reads

$$ 
{\partial \over \partial t} E(k,t) = \int \int_{\DC}T(k,p,q,t)dpdq, 
\eqno ({\rm A}9)
$$
where
$$ 
\eqalignno {
T(k,p,q,t) &\equiv {2 \over \pi k} \theta _{kpq}(t)(p^2-q^2)
\left [ {p^2-q^2\over pq}E(p,t)E(q,t) \right .  \cr
& \left . - {k^2-q^2\over kq}E(q,t)E(k,t) + {k^2-p^2\over kp} E(p,t)E(k,t) \right ]
\sin \alpha,   &({\rm A}10)  }
$$
and where $\nu _k \equiv \nu (k)k^2$, $\alpha $ is an angle opposite to the vector
${\bf k}$ in the triangle $\k + \p + \q = 0$, and the integration domain $\DC $ is 
defined by the triangular inequalities $ |k-p|<q<k+p$.  The function $\nu (k)$ given
by 
(A6) is used to compute the relaxation time $\theta _{kpq}(t) \equiv 
(1-e^{(\nu _k+\nu _p+ \nu _q)t})/(\nu _k +\nu _p +\nu _q)$ in (A10).

Following Kraichnan (1976) one can now define the two-parametric, or effective 
eddy viscosity at wave number $k$ in terms of the energy transfer
from all subgrid scale modes with $k > k_c$ to the given explicit mode $k$,
$$
\nu (k \vert k_c )= -T(k \vert k_c )/[2k^2E(k)],  \eqno({\rm A}11)
$$
where
$$
T(k \vert k_c ) \equiv \int \int ^{'} _{\DC }T(k,p,q)dpdq, \, ~k < k_c, 
\eqno ({\rm A}12)
$$
and where the integration is extended over all $p$ and $k$ such that $p$ 
and/or $q > k_c$.  Assuming that the limit $t \to \infty $ in (A10) 
is considered the time argument in T(k,p,q) has been omitted.  In that limit,
$\theta_{kpq} = (\nu _k +\nu _p +\nu _q)^{-1}$.

For $k \ll k_c$ the two-parametric viscosity $\nu (k \vert k_c )$
can be calculated analytically (Kraichnan, 1976). In this case, the triangular
inequality becomes $\vert p - q \vert \le k \ll q$. Therefore, all the 
quantities that enter $T(k,p,q)$ can be expanded in powers of $p - q$. Then, 
the $p$ integration can be performed resulting in
$$
\nu(k \vert k_c) = {1 \over 4}\int ^{\infty} _{\k_c}\theta _{kqq}
{d\over dq}[qE(q)]dq, ~~~~~k \ll k_c.     \eqno ({\rm A}13)
$$

Substitution of (A6) and (A8) into (A13) gives the asymptotic eddy viscosity 
for the largest scales,
$$
\nu(0 \vert k_c) = - {1 \over 3} \left ( {3 D_o \over 64 \pi} \right )^{1/3} 
k_c^{-4/3}.  \eqno ({\rm A}14)
$$

For arbitrary $k$, $\nu (k \vert k_c )$ was calculated via numerical integration of
(A10); the resulting normalized two-parametric viscosity $N(k/k_c)$, Eq. (6), is shown 
in Fig. 1. This function is negative for $k \ll k_c$ and positive for $k \to k_c$. 

Noting that in the energy transfer subrange the energy injection rate $\overline
\epsilon$
is equal to the rate of energy transfer from all the subgrid modes $k > k_c$ to all 
explicit modes $k < k_c$, one can find the relation between the forcing amplitude
$D_0$ 
in (A2) and $\overline \epsilon$.  Indeed, as follows from (A11) and (A12), 
$$
\overline \epsilon = - 2 \int_0^{k_c} \nu(k \vert k_c) E(k) k^2 dk.
\eqno ({\rm A}15)
$$
Substituting (A8) and (A14) into (A15) and performing numerical integration one
finds
$$
D_0 \simeq  63 \overline \epsilon.   \eqno ({\rm A}16)
$$
Using (A16) one can now calculate the Kolmogorov constant $C_K$ in (A8) and the
numerical factor in (A14), which are respectively, 
$$
E(k) = C_K {\overline \epsilon}^{2/3}k^{-5/3}, ~~~~~~C_K \simeq 5.12, 
\eqno ({\rm A}17)
$$
and
$$
\nu(0 \vert k_c) = - 0.327 {\overline \epsilon}^{1/3} k_c^{-4/3}.  
\eqno ({\rm A}18)
$$

These results have been used in Eqs. (8) and (14).

\head {9. References}

\item{} Batchelor, G.K.  (1969).
Computation of the energy spectrum in homogeneous two-dimensional
turbulence.  {\it Phys.~Fluids Suppl.~II}, {\bf 12}, 233--238.

\item{} Browning, G.L., and Kreiss, H.-O. (1989).
Comparison of numerical methods for the calculation of
two-dimensional turbulence. {\it Math.~Comput.}, {\bf 52}, 369--388.

\item{} Chekhlov, A., and Orszag, S. A., Sukoriansky, S., Galperin,
B., Staroselsky, I. (1994). Direct Numerical Simulation Tests of Eddy
Viscosity in Two Dimensions. {\it Phys. Fluids}, {\bf 6}, 2548--2550. 

\item{} Dannevik, W. P., Yakhot, V., and Orszag, S. A. (1987).
Analytical Theories of Turbulence and the $\epsilon$-Expansion. {\it
Phys. Fluids}, {\bf 30}, 2021--2029. 

\item{} Domaradzki, J. A., Metcalfe R. W., Rogallo, R. S., and
Riley J. J. (1987). Analysis of Subgrid-Scale Eddy Viscosity with Use
of Results from Direct Numerical Simulations. {\it Phys. Rev. Lett.},
{\bf 58}, 547--550. 

\item{} Forster, D., Nelson, D.R., and Stephen, M.J. (1977).
Large distance and long-time properties of a randomly stirred fluid.
{\it Phys.~Rev.~A}, {\bf 16}, 732--749.

\item{} Galperin, B., and Orszag, S.A. (ed.),
{\it Large Eddy Simulation of Complex Engineering and Geophysical
Flows}, Cambridge University Press. 

\item{} Gama, S., Frisch, U., and Scholl, H. (1991). The
Two-Dimensional Navier-Stokes Equations with a Large-Scale Instability
of the Kuramoto-Sivashinsky Type: Numerical Exploration on the
Connection Machine. {\it J. Sci. Comput.}, {\bf 6}, 425--452. 

\item{} Gama, S., Vergassola, M., and Frisch, U. (1994). Negative
Eddy Viscosity in Isotropically Forced Two-Dimensional Flow: Linear
and Nonlinear Dynamics. {\it J. Fluid Mech.}, {\bf 260}, 95--125. 

\item{} Karniadakis, G. E., Israeli, M., and Orszag, S. A. (1991). High-Order
Splitting Methods for the Incompressible Navier-Stokes Equations. 
{\it J. Comp.  Phys.}, {\bf 97}, 414--443. 

\item{} Kraichnan, R.H. (1967).
Inertial ranges in two-dimensional turbulence.
{\it Phys.~Fluids}, {\bf 10}, 1417--1423.

\item{} Kraichnan, R. H. (1971). Inertial-Range Transfer in Two- and
Three-Dimensional Turbulence, {\it J. Fluid Mech.}, {\bf 47}, 525--535. 

\item{} Kraichnan, R. H. (1976). Eddy Viscosity in Two and
Three Dimensions. {\it J. Atmos. Sci.}, {\bf 33}, 1521--1536. 

\item{} Kraichnan, R. H., and Montgomery, D. (1980). Two-dimensional
turbulence. {\it Rep. Prog. Phys.}, {\bf 43}, 547--619. 
\item{} Kuramoto, Y., (1984).
{\it Chemical Oscillations, Waves, and Turbulence}.  Springer-Verlag.

\item{} Kuramoto, Y., and Tsuzuki, T. (1976). Persistent
Propagation of Concentration waves in Dissipative Media Far from
Thermal Equilibrium. {\it Progr. Theoret. Phys.}, {\bf 55}, 365--369. 

\item{} Lesieur, M., (1990).
{\it Turbulence in Fluids}, 2nd Edition,  Kluwer.

\item{} Monin, A. S., (1986). 
Stratification and circulation of the ocean.
In: {\it Synoptic Eddies in the Ocean}. Eds. V.M. Kamenkovich,
M.N. Koshlyakov, and A.S. Monin, D. Riedel Publishing Company, pp. 1--33.

\item{} Monin, A. S., and Yaglom, A. M., (1975). {\it Statistical Fluid
Mechanics}. MIT Press, Cambridge. 

\item{} Orszag, S. A. (1969). High-Speed Computing in Fluid
Dynamics. {\it Phys. Fluids}, {\bf Suppl. II}, II250--II257. 

\item{} Sadourny, R., and Basdevant, C., (1985).
Parameterization of subgrid scale barotropic and baroclinic eddies
in quasi-geostrophic models:  Anticipated potential vorticity method.
{\it J.~Atmos.~Sci.}, {\bf 42}, 1353--1363.

\item{} Sivashinsky, G. I. (1977). Nonlinear Analysis of
Hydrodynamic Instability in Laminar Flames-I. Derivation of Basic
Equations. {\it Acta Astr.}, {\bf 4}, 1177--1206. 

\item{} Smagorinsky, J. (1963). General Circulation Experiments With
The Primitive Equations. {\it Mont. Weather Rev.}, {\bf 91}, 99--164. 

\item{} Smagorinsky, J. (1993). Some Historical Remarks on the Use
of Nonlinear Viscosities. In Galperin, B., and Orszag, S. A. (ed.),
{\it Large Eddy Simulation of Complex Engineering and Geophysical
Flows}, Cambridge University Press, New York, pp. 3--36. 

\item{} Smith, L., and Yakhot, V. (1993a). Bose Condensation and
Small-Scale Structure Generation in a Random Force Driven 2D
Turbulence. {\it Phys. Rev. Lett.}, {\bf 71}, 352--355. 

\item{} Smith, L., and Yakhot, V. (1993b). Finite-Size Effects in Forced 
Two-Dimensional Turbulence. {\it J. Fluid Mech.}, {\bf 274}, 115--138. 

\item{} Sommeria, J. and Moreau, R., (1982). Why, How and When, MHD
Turbulence Becomes Two-Dimensional. {\it J.~Fluid~Mech.}, {\bf 118}, 507--518.

\item{} Staroselsky, I., and Sukoriansky, S. (1993).
Renormalization Group Approach to Two-Dimensional Turbulence and the
$\epsilon$-Expansion for the Vorticity Equation. In Branover, H., and
Unger, Y. (ed.), {\it Advances in Turbulence Studies}, Progress in
Astron. and Aeron., AIAA, {\bf 149}, 159--164. 

\item{} Staroselsky, I., Sukoriansky, S., and Orszag, S.A.  (1995).
The $\epsilon-$expansion procedure for near-equilibrium two-dimensional 
randomly stirred fluids.  {\it Phys. Rev. E}, submitted.

\item{} Starr, V. P., (1968). {\it Physics of Negative Viscosity Phenomena}, 
McGraw-Hill. 

\item{} Yakhot, A., Orszag, S. A., Yakhot, V., and Israeli, M.
(1989). Renormalization Group Formulation of Large-Eddy Simulations.
{\it J. Sci. Comput.}, {\bf 4}, 139--158. 

\item{} Yakhot, V., and Orszag, S. A. (1986). Renormalization Group
Analysis of Turbulence. I. Basic Theory. {\it J. Sci. Comput.}, {\bf 1}, 3--51. 

\item{} Yakhot, V., and Sivashinsky, G. I., (1985). Negative viscosity effect 
in large-scale flows. {\it Phys. Fluids}, {\bf 28}, 1040--1042. 

\head {9. List of the Figure Captions}

\item{Figure 1.}  Normalized two-parametric eddy viscosity from DNS (dots), 
from TFM (dashed line), and from RG (solid line) (from Chekhlov et al., 
1994).

\item{Figure 2.}  Actual two-parametric eddy viscosity from DNS (dots) 
and from RG (solid line).  In RG calculations, the energy spectrum for 
$k < 5$ was corrected in accordance with the DNS data (from Chekhlov et al., 
1994).

\item{Figure 3.}  The evolution of the total energy ${\overline  E}(t)$ (a)
and total enstrophy ${\overline \om}(t)$ in Case 1 LES.  Figure 3a also shows
the evolution of ${\overline  E}(t)$ with the energy of the 1st, 2nd, 3rd,
4th, 5th, 6th and 7th modes removed.

\item{Figure 4.}  The evolution of the instantaneous energy spectrum for
$t/\tau_{tu} = 0.56, 1.11, 1.67 and 2.78$.  The solid line shows the 
Kolmogorov -5/3 slope.

\item{Figure 5.}  Same as Fig. 3 but for Case 2 LES.
\item{Figure 6.}  Same as Fig. 4 but for Case 2 LES.  Note that after
$t/\tau_{tu} \simeq 2$ all instantaneous profiles $E(k,t)$ become close
to Kolmogorov law (9).

\item{Figure 7.}  Same as Fig. 3 but for Case 3 LES.  Because the
amplitudes of the first four modes are set to zero, only the evolution
of ${\overline  E}(t)$ with the energy of the 4th, 5th, 6th and 7th modes 
removed is shown.

\item{Figure 8a.}  Same as Fig. 6 but for Case 3 LES.  Note that since the
Kolmogorov scaling is attained after about $t/\tau_{tu} = 2$, only time average
energy spectrum is shown.

\item{Figure 8b.}  The energy flux $\Pi_E(k)$ for Case 3 LES. 

\item{Figure 9.}  Same as Fig. 7 but for Case 4 LES.  

\item{Figure 10a.}  Same as Fig. 8 but for Case 4 LES. 

\item{Figure 10b.}  The time averaged Kolmogorov constant $C_K$ for Case 4 LES.

\item{Figure 10c.}  The energy flux $\Pi_E(k)$ for Case 4 LES. 

\item{Figure 11.}  Same as Fig. 7 but for Case 5 LES.

\item{Figure 12.}  Same as Fig. 8 but for Case 5 LES. 

\item{Figure 13.}  Same as Fig. 7 but for Case 6 LES.

\item{Figure 14a.}  Same as Fig. 8 but for Case 6 LES. 

\item{Figure 14b.}  The time averaged Kolmogorov constant $C_K$ for Case 6 LES.

\end
\vfill\bye

%% file: djnlx.tex


\def\undertext#1{$\underline{\smash{\hbox{#1}}}$}


  \font\twelverm=cmr12		         \font\twelvei=cmmi12
  \font\twelvesy=cmsy10 scaled 1200      \font\twelveex=cmex10 scaled 1200
  \font\twelvebf=cmbx12			 \font\twelvesl=cmsl12
  \font\twelvett=cmtt12		         \font\twelveit=cmti12

  \font\twelvemib=cmmib10 scaled 1200 
  \font\elevenmib=cmmib10 scaled 1095
  \font\tenmib=cmmib10
  \font\eightmib=cmmib10 scaled 800


\font\elevenrm=cmr10 scaled 1095    \font\eleveni=cmmi10 scaled 1095
\font\elevensy=cmsy10 scaled 1095



\skewchar\eleveni='177   \skewchar\elevensy='60
\skewchar\elevenmib='177

\newfam\mibfam%


  \skewchar\twelvei='177   \skewchar\twelvesy='60
  \skewchar\twelvemib='177
%
%
\def\twelvepoint{\normalbaselineskip=12.4pt
  \abovedisplayskip 12.4pt plus 3pt minus 9pt
  \belowdisplayskip 12.4pt plus 3pt minus 9pt
  \abovedisplayshortskip 0pt plus 3pt
  \belowdisplayshortskip 7.2pt plus 3pt minus 4pt
  \smallskipamount=3.6pt plus 1.2pt minus 1.2pt
  \medskipamount=7.2pt plus 2.4pt minus 2.4pt
  \bigskipamount=14.4pt plus 4.8pt minus 4.8pt
  \def\rm{\fam0\twelverm}          \def\it{\fam\itfam\twelveit}%
  \def\sl{\fam\slfam\twelvesl}     \def\bf{\fam\bffam\twelvebf}%
  \def\mit{\fam 1}                 \def\cal{\fam 2}%
  \def\tt{\twelvett}%
  \def\mib{\fam\mibfam\twelvemib}%
 
  \textfont0=\twelverm   \scriptfont0=\tenrm     \scriptscriptfont0=\sevenrm
  \textfont1=\twelvei    \scriptfont1=\teni      \scriptscriptfont1=\seveni
  \textfont2=\twelvesy   \scriptfont2=\tensy     \scriptscriptfont2=\sevensy
  \textfont3=\twelveex   \scriptfont3=\twelveex  \scriptscriptfont3=\twelveex
  \textfont\itfam=\twelveit 
  \textfont\slfam=\twelvesl
  \textfont\bffam=\twelvebf 
  \textfont\mibfam=\twelvemib       \scriptfont\mibfam=\tenmib
                               	      \scriptscriptfont\mibfam=\eightmib

  \def\xrm{\textfont0=\twelverm\scriptfont0=\tenrm
      \scriptscriptfont0=\sevenrm\rm}
\normalbaselines\rm}


\mathchardef\alpha="710B
\mathchardef\beta="710C
\mathchardef\gamma="710D
\mathchardef\delta="710E
\mathchardef\epsilon="710F
\mathchardef\zeta="7110
\mathchardef\eta="7111
\mathchardef\theta="7112
\mathchardef\kappa="7114
\mathchardef\lambda="7115
\mathchardef\mu="7116
\mathchardef\nu="7117
\mathchardef\xi="7118
\mathchardef\pi="7119
\mathchardef\rho="711A
\mathchardef\sigma="711B
\mathchardef\tau="711C
\mathchardef\phi="711E
\mathchardef\chi="711F
\mathchardef\psi="7120
\mathchardef\omega="7121
\mathchardef\varepsilon="7122
\mathchardef\vartheta="7123
\mathchardef\varrho="7125
\mathchardef\varphi="7127



\def\beginlinemode{\endmode
  \begingroup\parskip=0pt \obeylines\def\\{\par}\def\endmode{\par\endgroup}}
\def\beginparmode{\endmode
  \begingroup \def\endmode{\par\endgroup}}
\let\endmode=\par
{\obeylines\gdef\
{}}
\def\singlespace{\baselineskip=\normalbaselineskip}

\def\oneandahalfspace{\baselineskip=\normalbaselineskip
  \multiply\baselineskip by 3 \divide\baselineskip by 2}
\def\doublespace{\baselineskip=\normalbaselineskip \multiply\baselineskip by 2}

\nopagenumbers
\newcount\firstpageno
\firstpageno=2

\headline={\ifnum\pageno<\firstpageno{\hfil}\else{\hfil\elevenrm\folio\hfil}\fi}

\let\rawfootnote=\footnote		
\def\footnote#1#2{{\singlespace\parindent=0pt
\rawfootnote{$^{#1}$}{{\tenrm #2}}}}  
\def\raggedcenter{\leftskip=4em plus 12em \rightskip=\leftskip
  \parindent=0pt \parfillskip=0pt \spaceskip=.3333em \xspaceskip=.5em
  \pretolerance=9999 \tolerance=9999
  \hyphenpenalty=9999 \exhyphenpenalty=9999 }
\def\dateline{\rightline{\ifcase\month\or
  January\or February\or March\or April\or May\or June\or
  July\or August\or September\or October\or November\or December\fi
  \space\number\year}}
\def\received{\vskip 3pt plus 0.2fill
 \centerline{\sl (Received\space\ifcase\month\or
  January\or February\or March\or April\or May\or June\or
  July\or August\or September\or October\or November\or December\fi
  \qquad, \number\year)}}


\hsize=6.5truein
\hoffset=0truein
\vsize=8.9truein
\voffset=0truein
\hfuzz=0.1pt
\vfuzz=0.1pt
\parskip=\medskipamount
\twelvepoint		
\doublespace		
\overfullrule=0pt	



\def\title			
  {\null\vskip 3pt plus 0.2fill
   \beginlinemode \doublespace \raggedcenter \bf}

\def\author			
  {\vskip 3pt plus 0.2fill \beginlinemode
   \singlespace \raggedcenter}

\def\affil			
  {\vskip 3pt plus 0.1fill \beginlinemode
   \oneandahalfspace \raggedcenter \sl}

\def\abstract		
  {\vskip 3pt plus 0.3fill \beginparmode
    \narrower ABSTRACT: }

\def\endtitlepage		
  {\endpage			
   \body}

\def\body			
  {\beginparmode}		

\def\head#1{			
  \filbreak\vskip 0.5truein	
  {\immediate\write16{#1}
    \raggedcenter \uppercase{#1}\par}
   \nobreak\vskip 0.25truein\nobreak}

\def\subhead#1{
     \vskip 0.25truein
      \immediate\write16{#1}
       \leftline{\undertext{\bf{#1}}}
        \nobreak}

\def\refto#1{$^{#1}$}		

\def\references			
  {\head{References}		
   \beginparmode
   \frenchspacing \parindent=0pt \leftskip=1truecm
   \parskip=8pt plus 3pt \everypar{\hangindent=\parindent}}

\gdef\refis#1{\indent\hbox to 0pt{\hss[#1]~}}	

\gdef\journal#1, #2, #3, 1#4#5#6{		
    {\sl #1~}{\bf #2}, #3 (1#4#5#6)}		

\def\refstylenp{		
  \gdef\refto##1{ [##1]}				
  \gdef\refis##1{\indent\hbox to 0pt{\hss##1)~}}	
  \gdef\journal##1, ##2, ##3, ##4 {			
     {\sl ##1~}{\bf ##2~}(##3) ##4 }}

\def\refstyleprnp{		
  \gdef\refto##1{ [##1]}				
  \gdef\refis##1{\indent\hbox to 0pt{\hss##1)~}}	
  \gdef\journal##1, ##2, ##3, 1##4##5##6{		
    {\sl ##1~}{\bf ##2~}(1##4##5##6) ##3}}

\def\figurecaptions		
  {\endpage
   \beginparmode
   \head{Figure Captions}
}

\def\endpage			
  {\vfill\eject}

\def\endpaper			
  {\endmode\vfill\supereject}


\def\ref#1{Ref.[#1]}			

\def\frac#1#2{{\textstyle{#1 \over #2}}}

\def\etal{{\it et al. \/}}

\def\sla{\raise.15ex\hbox{$/$}\kern-.57em}
\def\leaderfill{\leaders\hbox to 1em{\hss.\hss}\hfill}
\def\twiddle{\lower.9ex\rlap{$\kern-.1em\scriptstyle\sim$}}
\def\bigtwiddle{\lower1.ex\rlap{$\sim$}}
\def\gtwid{\mathrel{\raise.3ex\hbox{$>$\kern-.75em\lower1ex\hbox{$\sim$}}}}
\def\ltwid{\mathrel{\raise.3ex\hbox{$<$\kern-.75em\lower1ex\hbox{$\sim$}}}}
\def\square{\kern1pt\vbox{\hrule height 1.2pt\hbox{\vrule width 1.2pt\hskip 3pt
   \vbox{\vskip 6pt}\hskip 3pt\vrule width 0.6pt}\hrule height 0.6pt}\kern1pt}
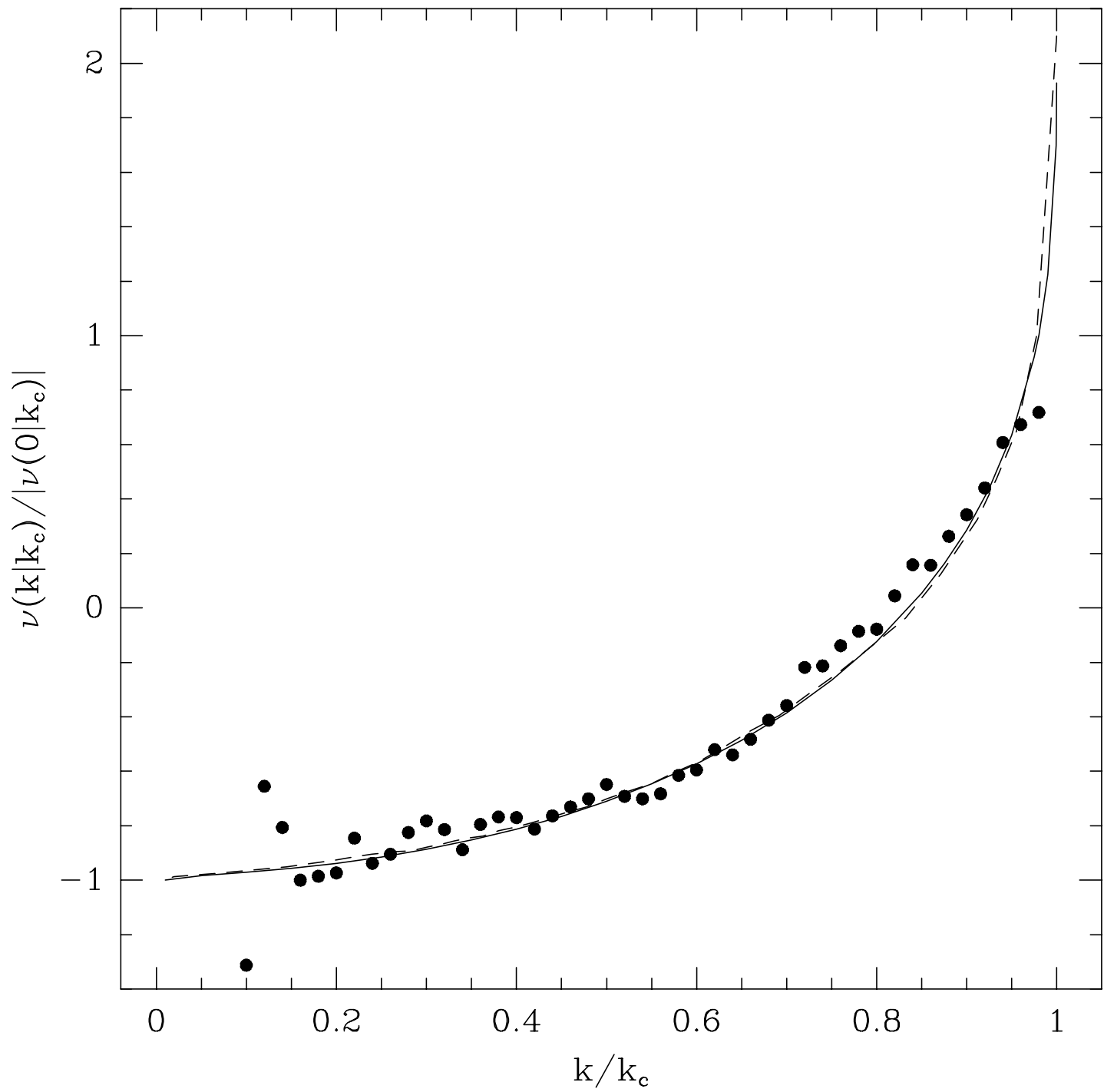
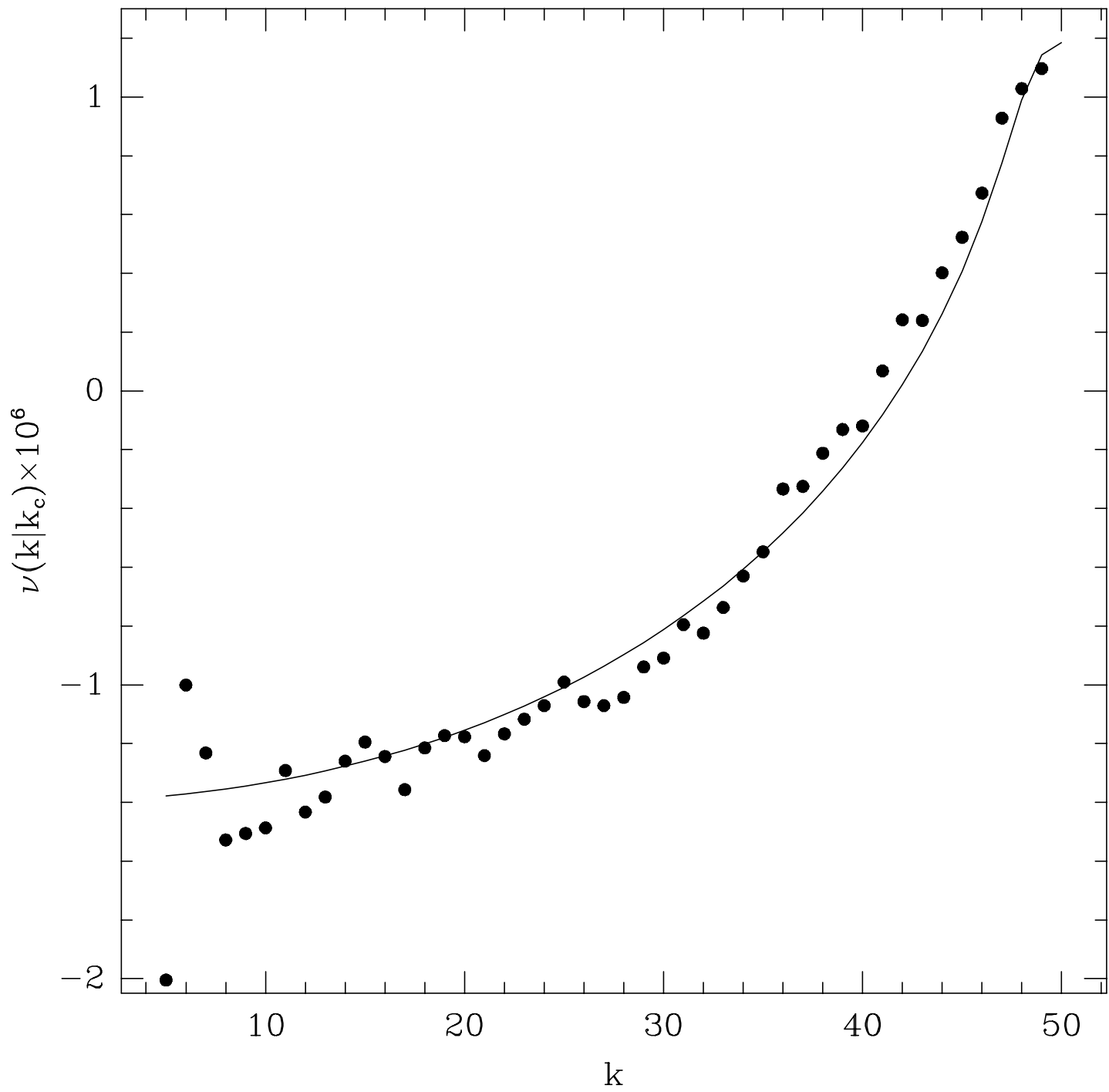
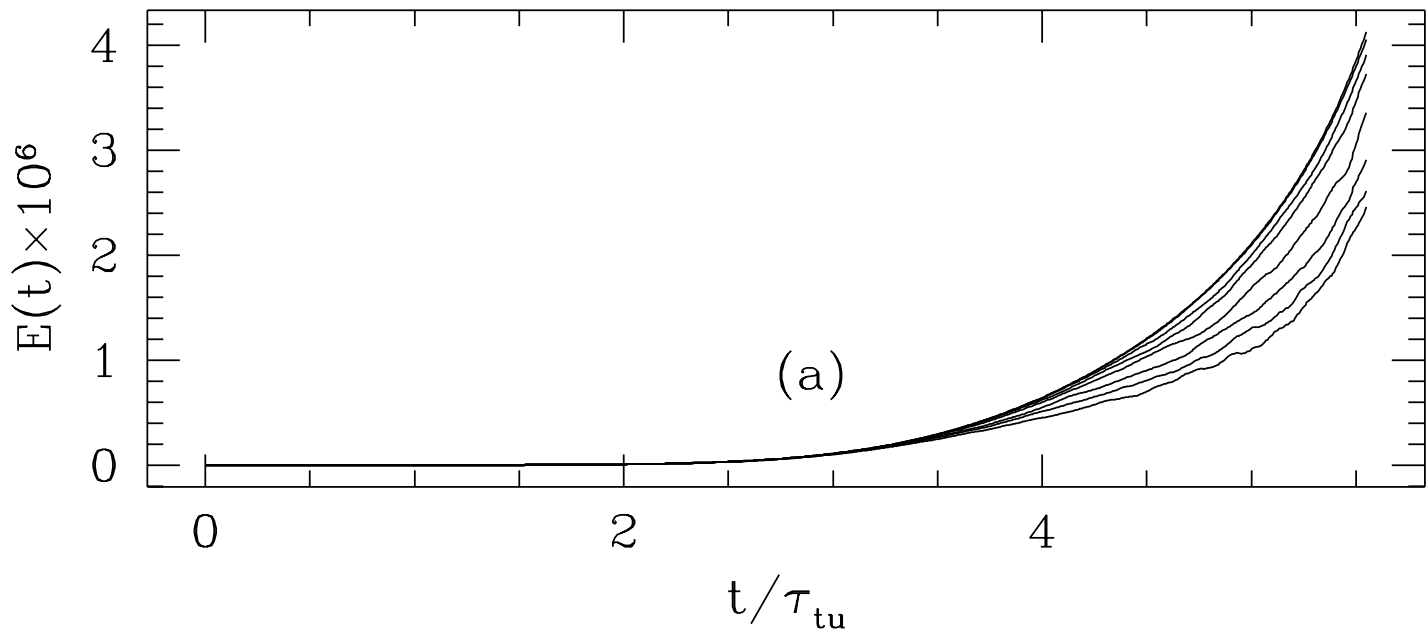
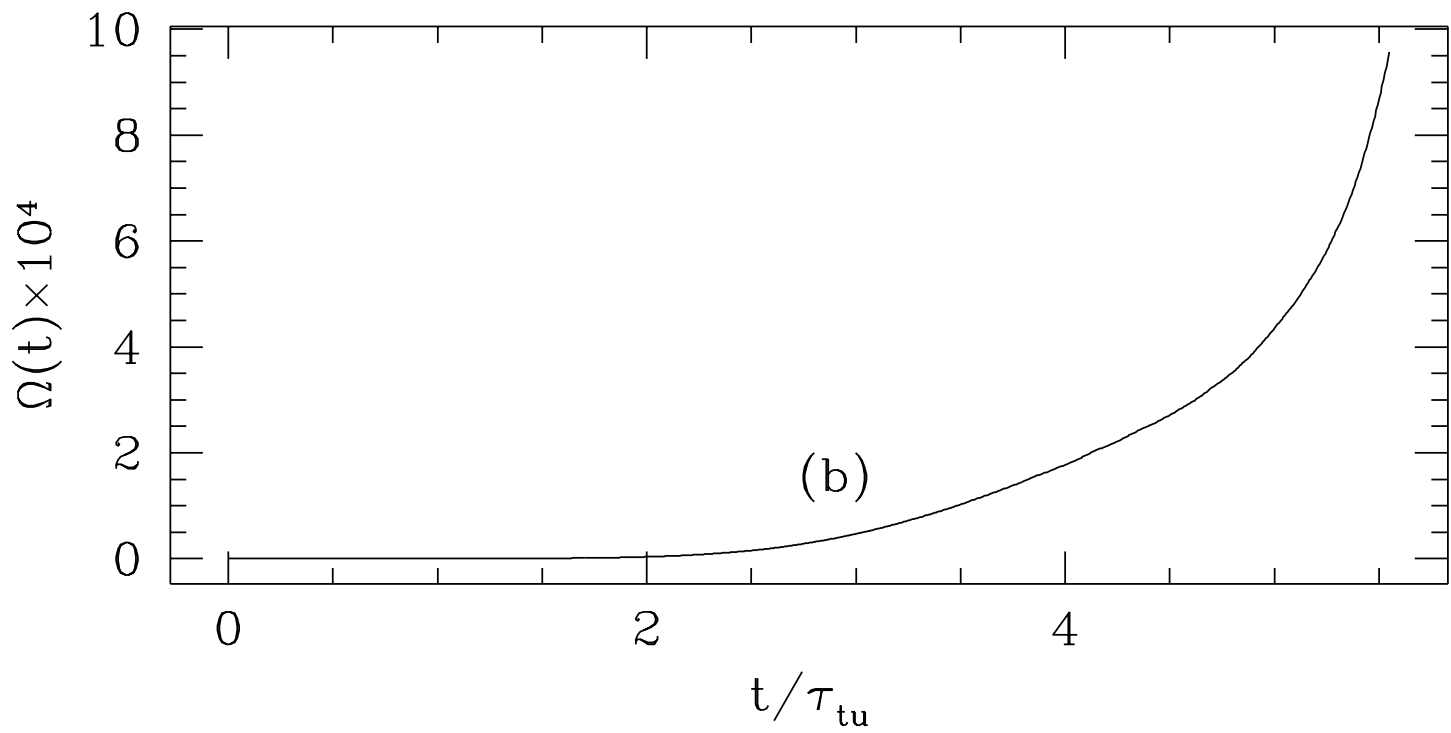
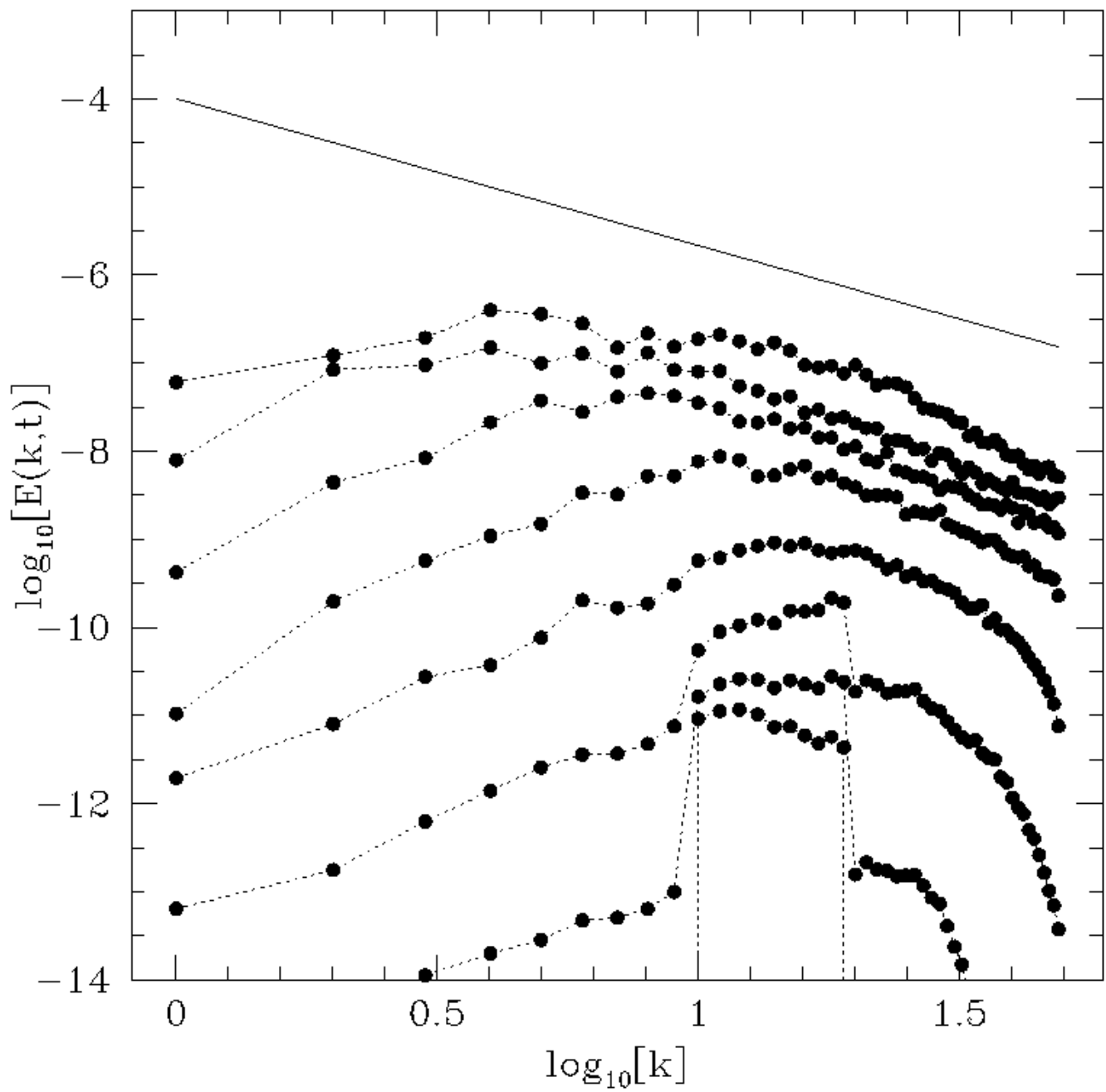
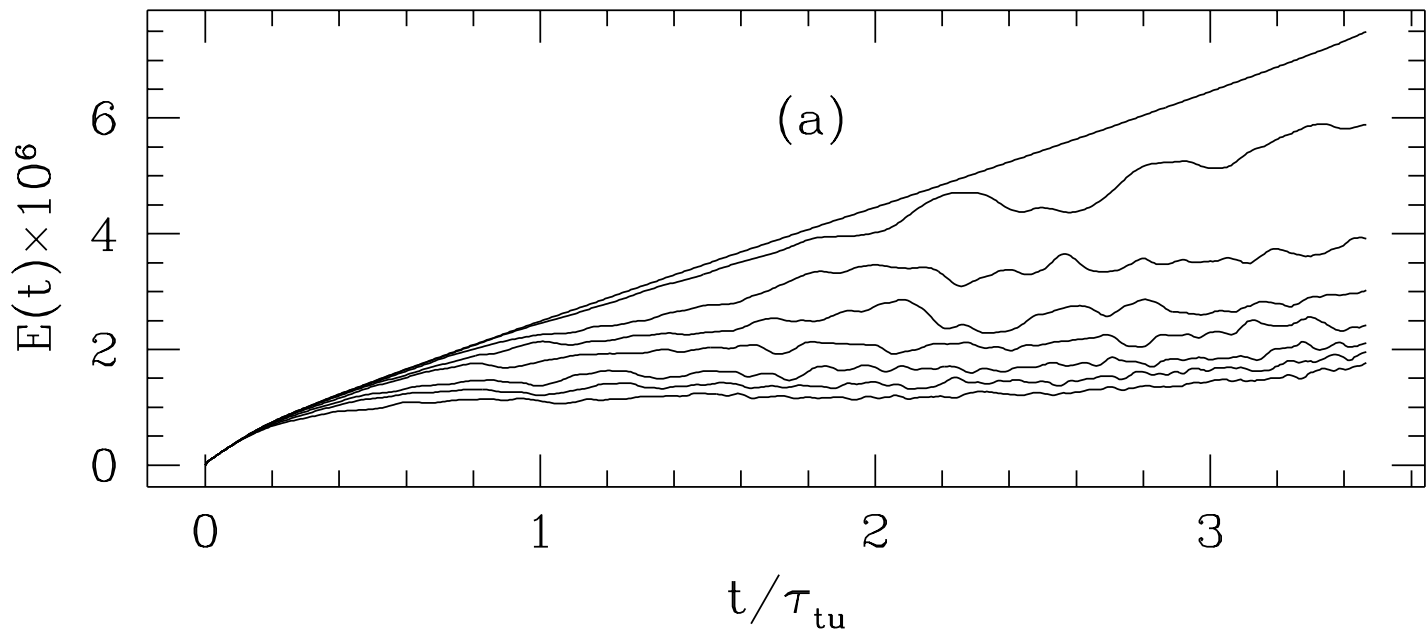
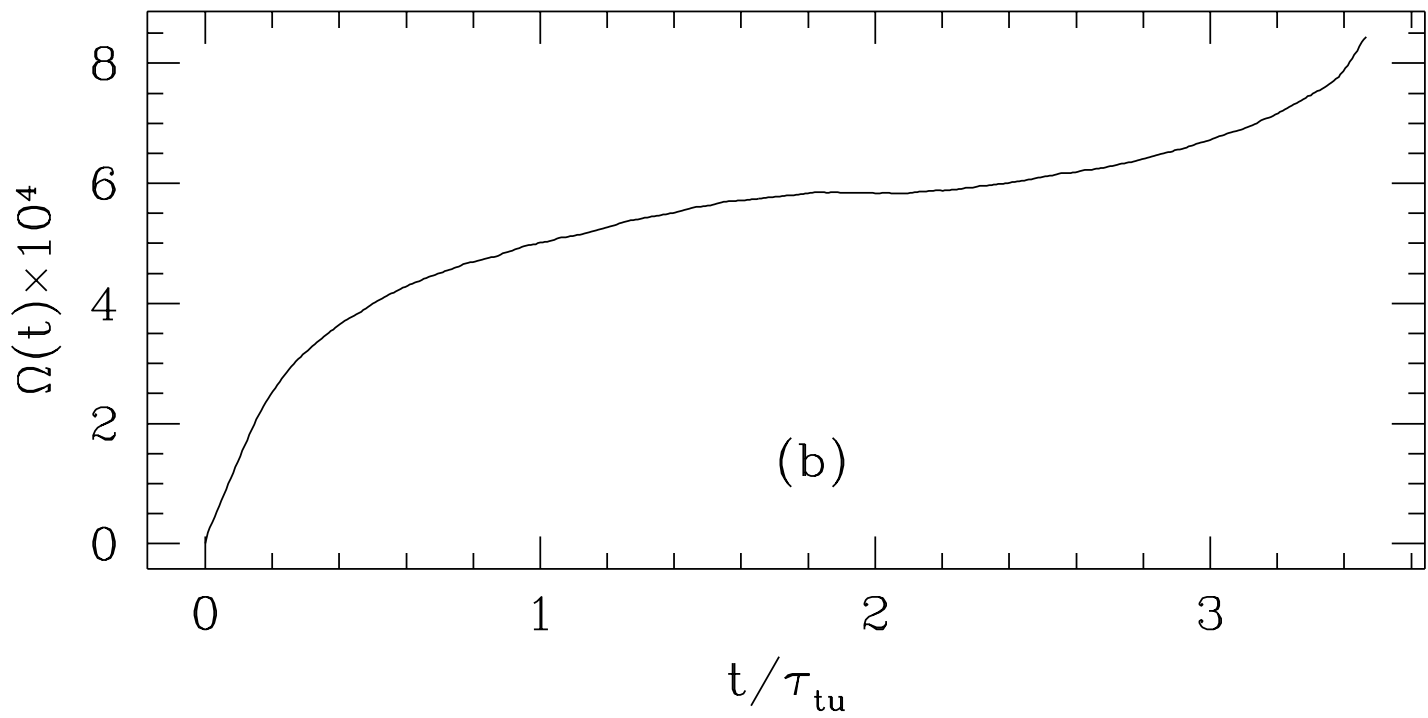
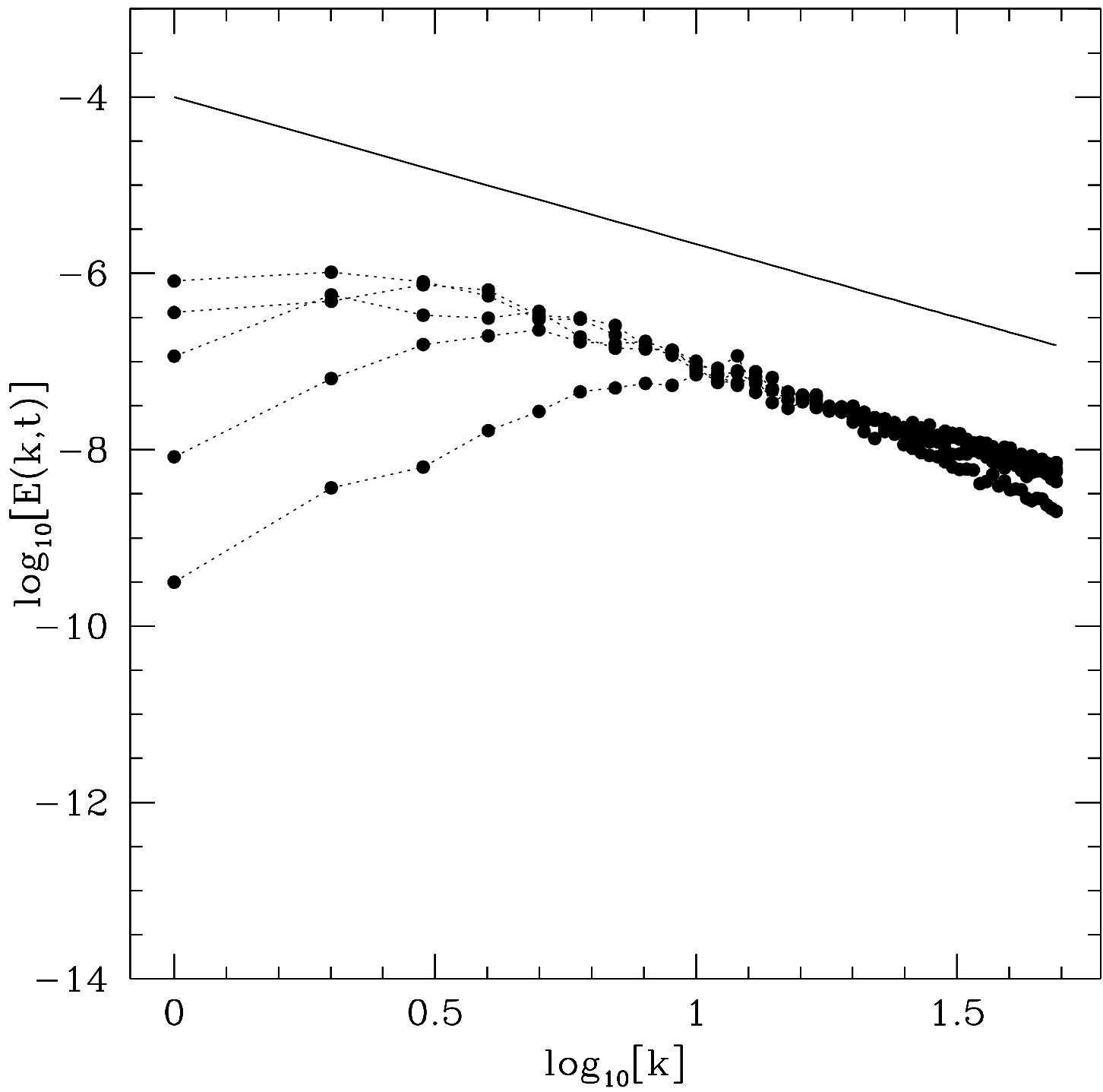
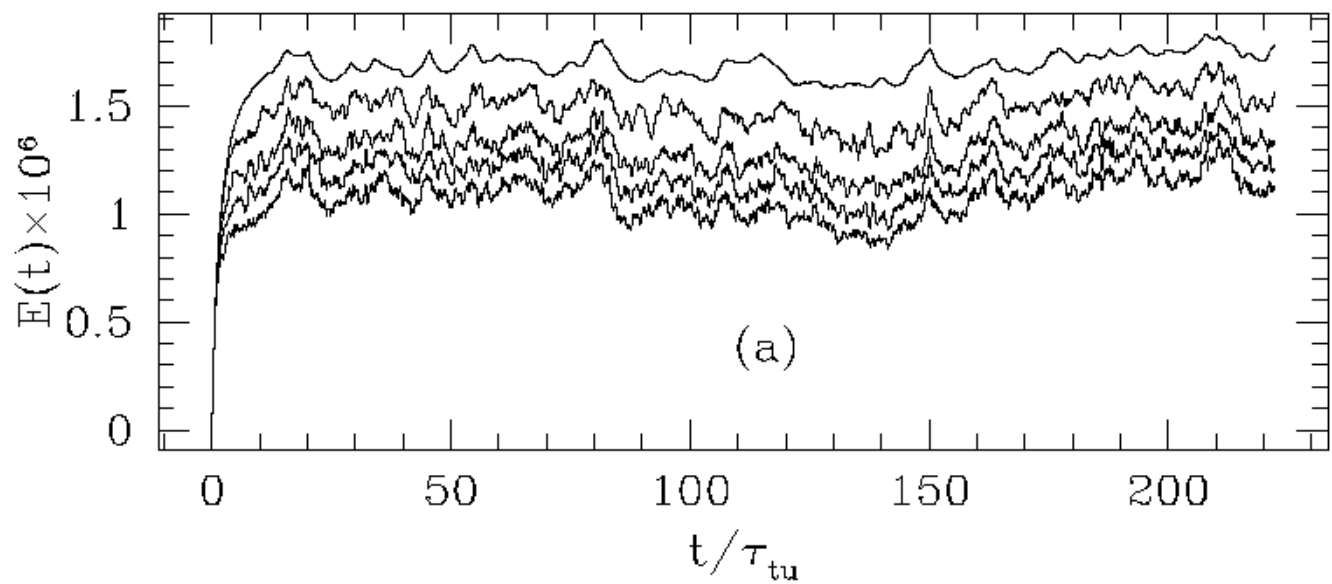
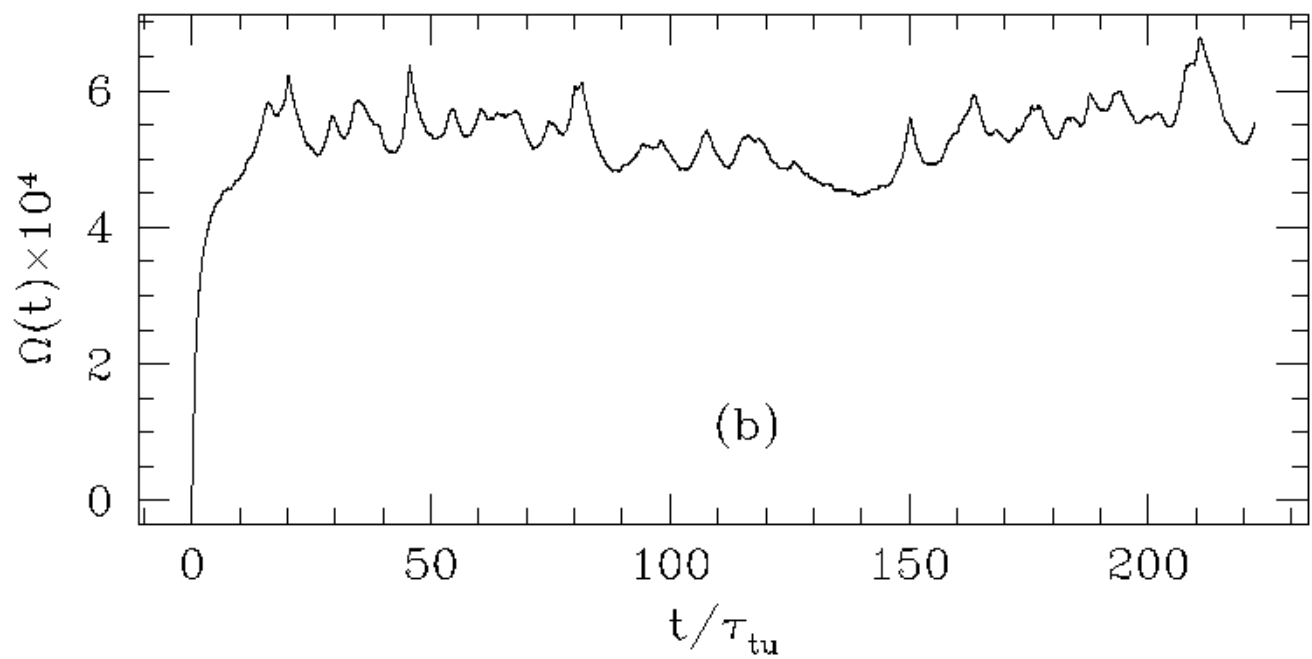
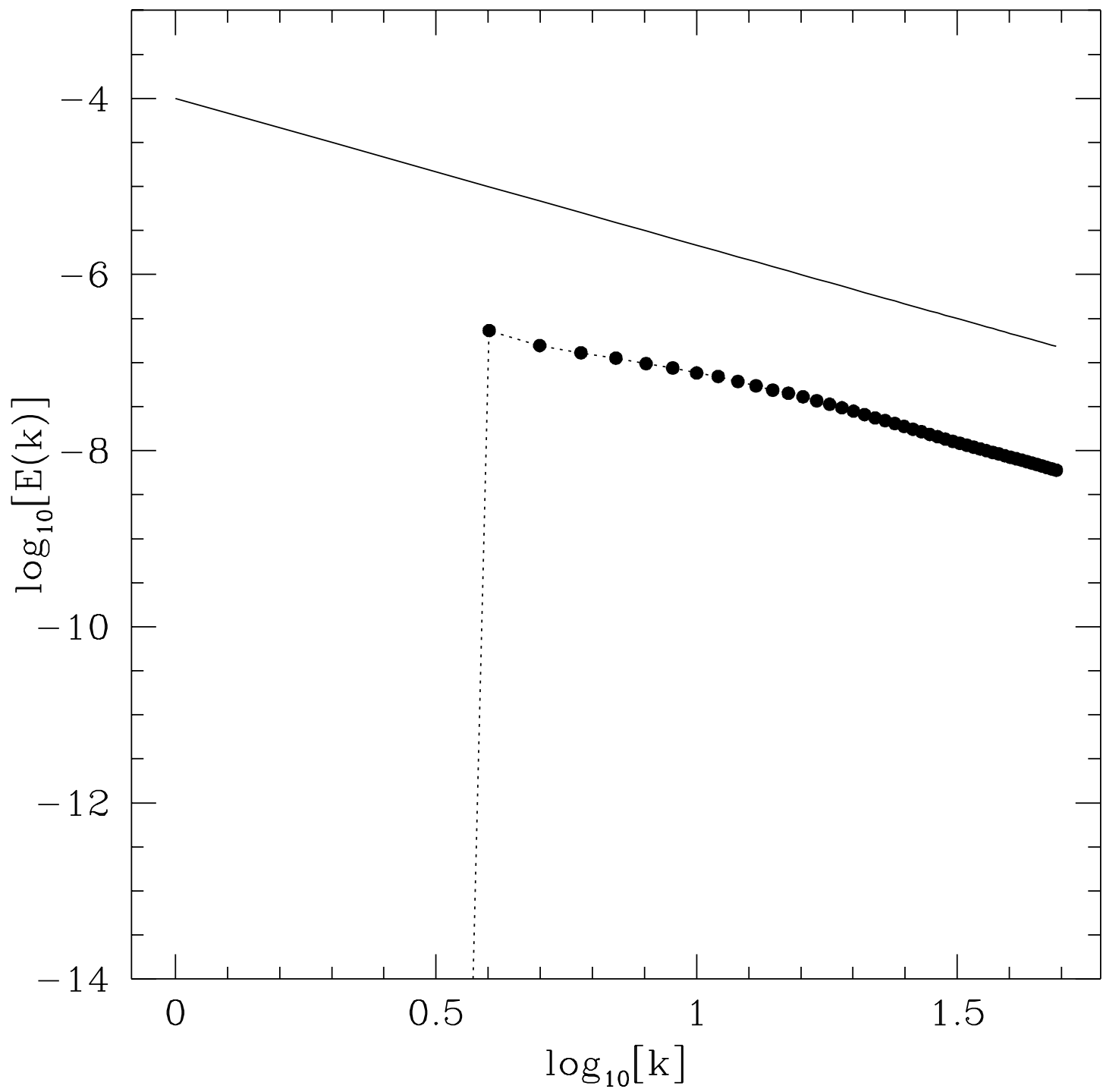
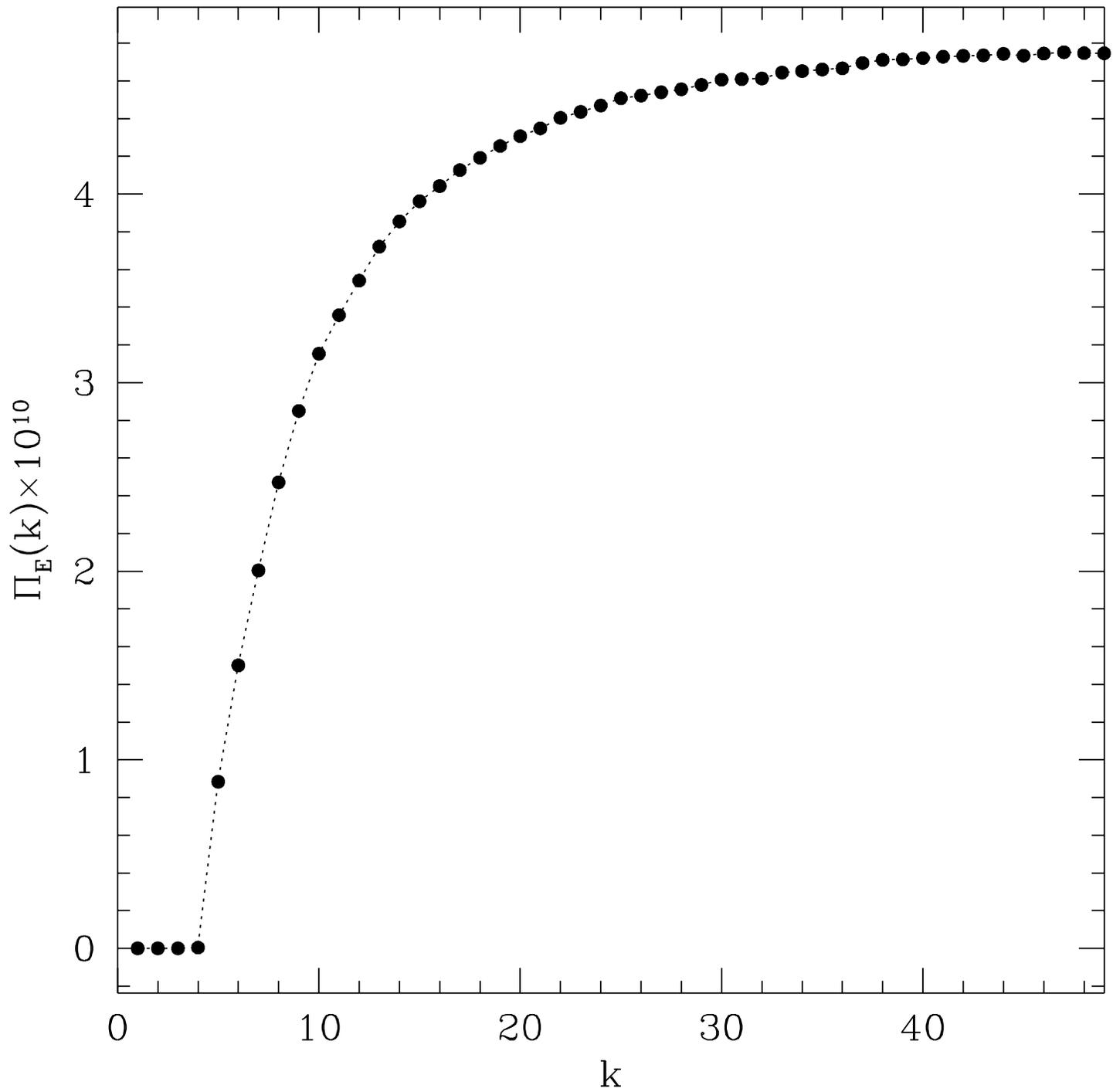
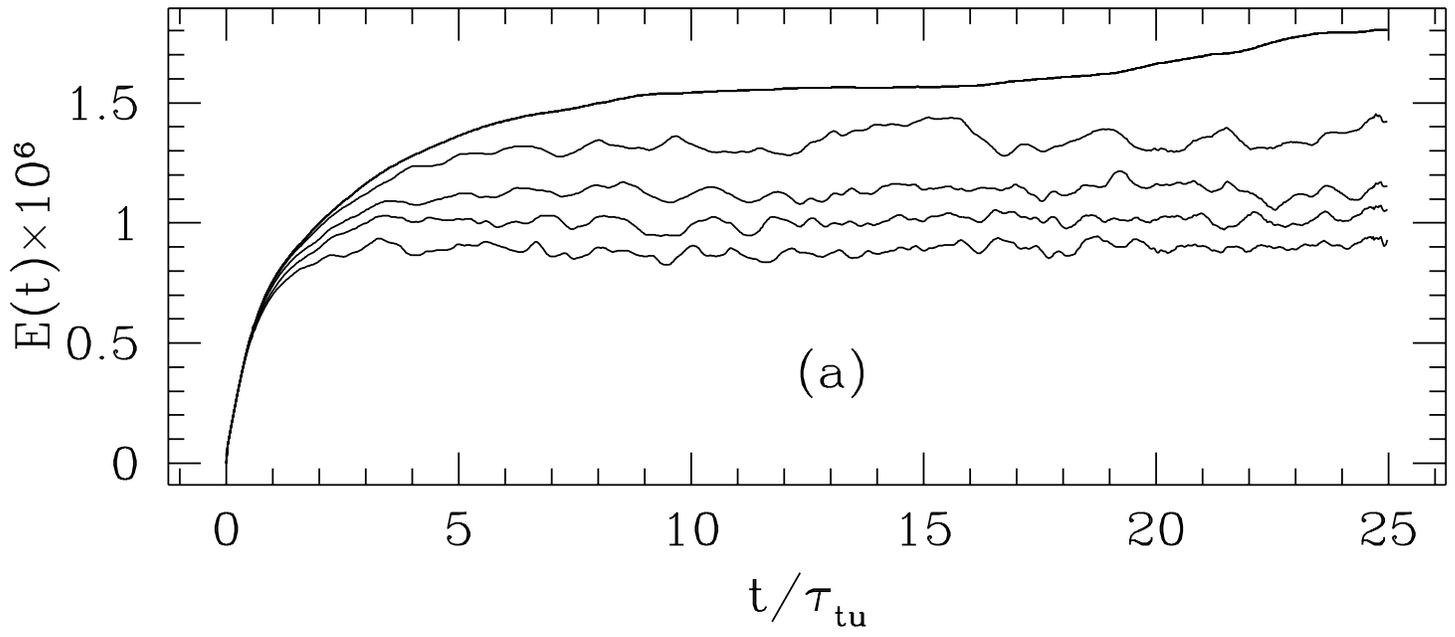
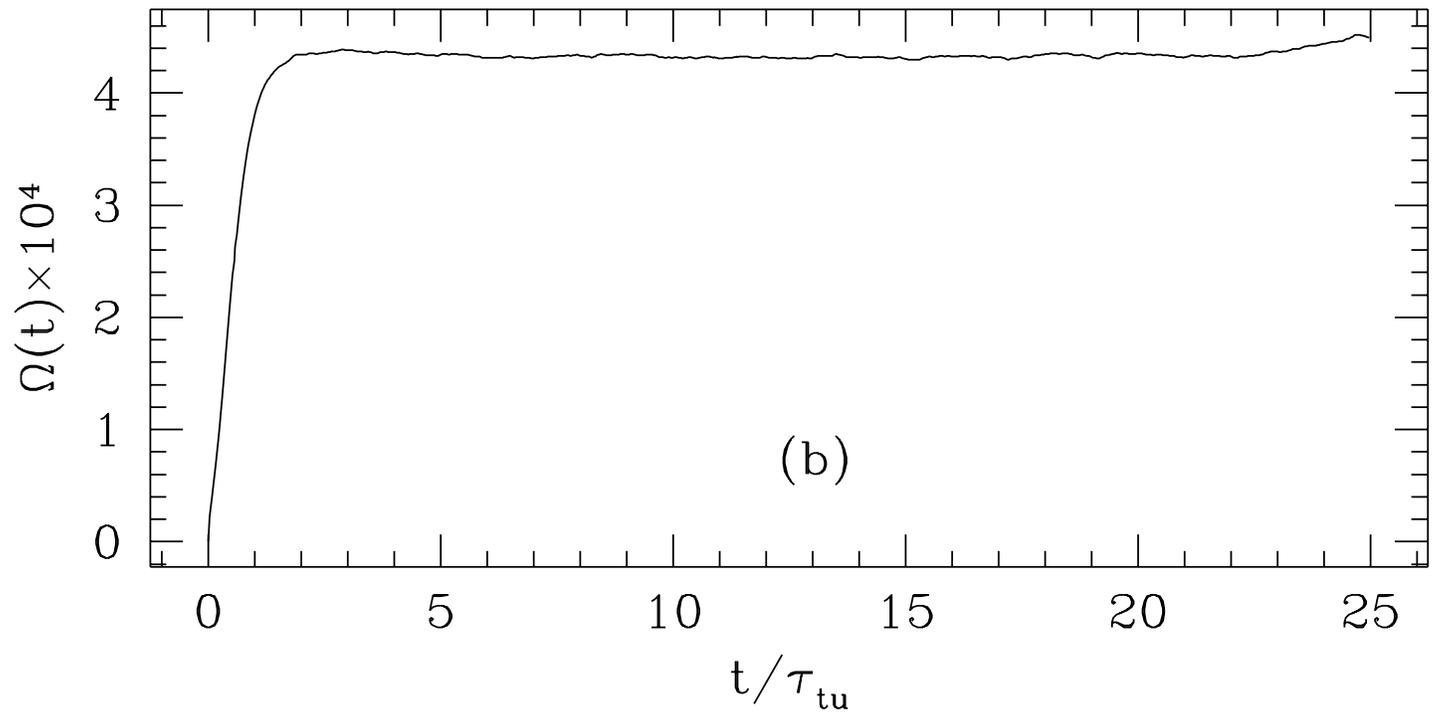
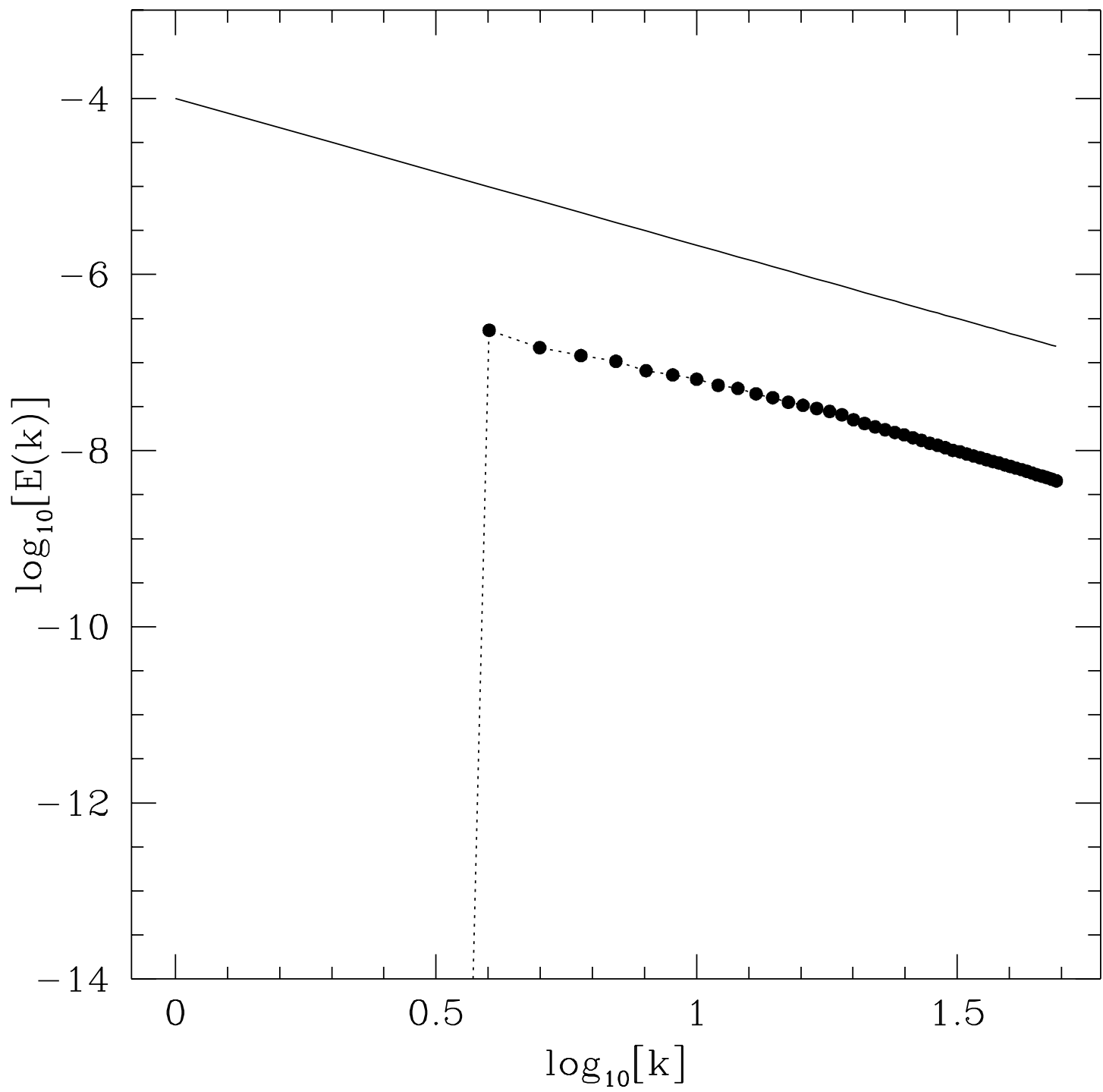
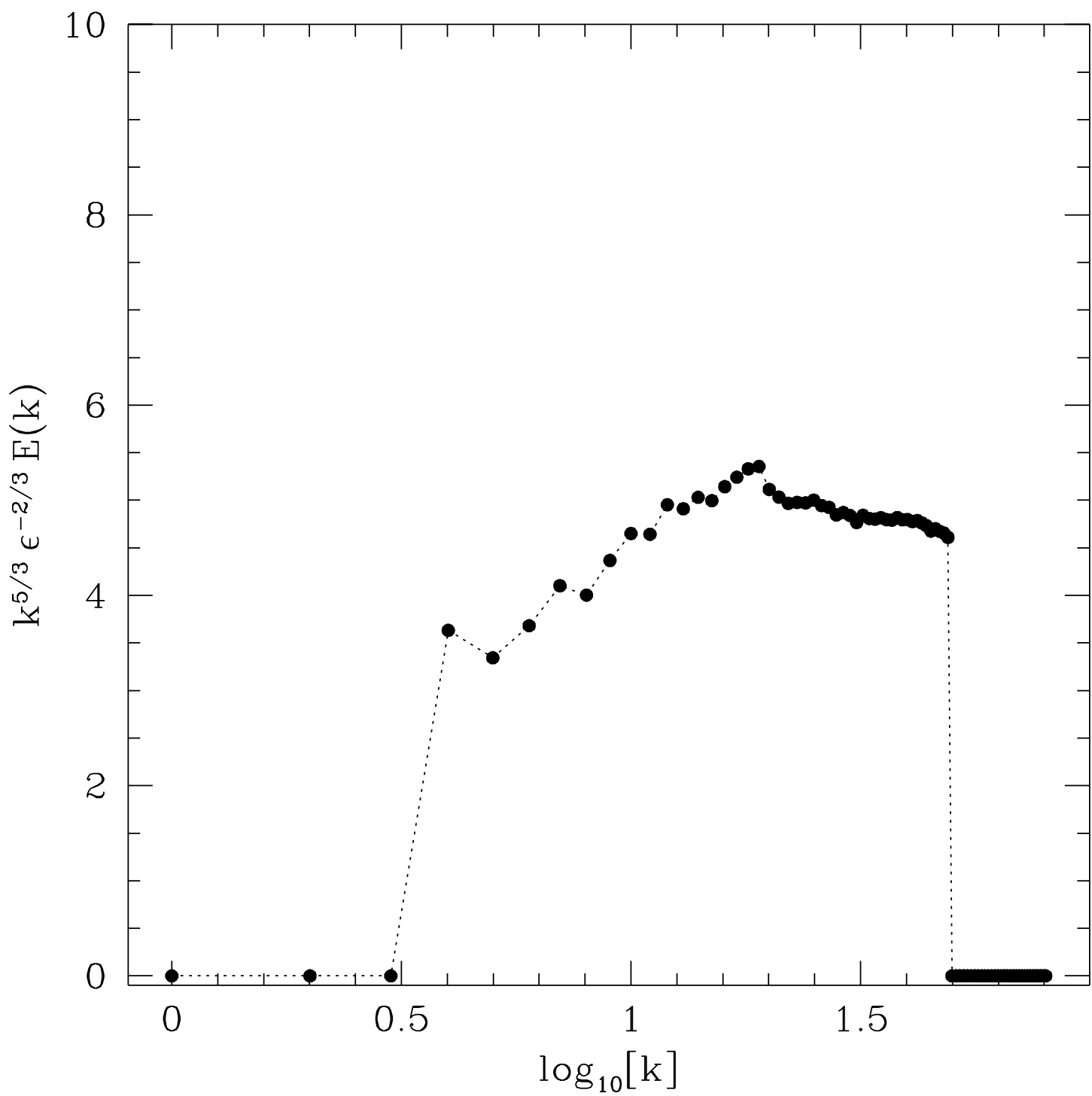
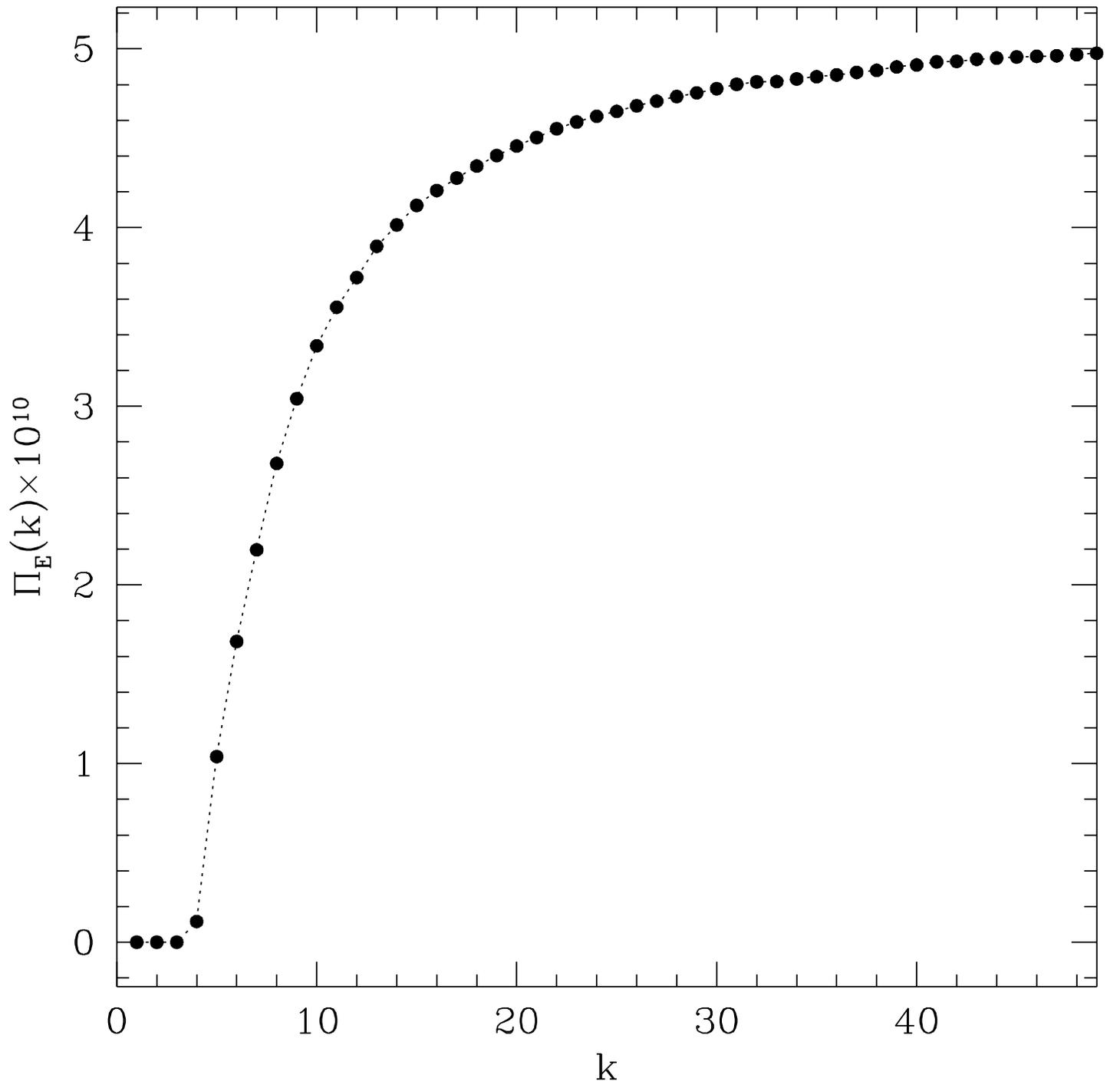
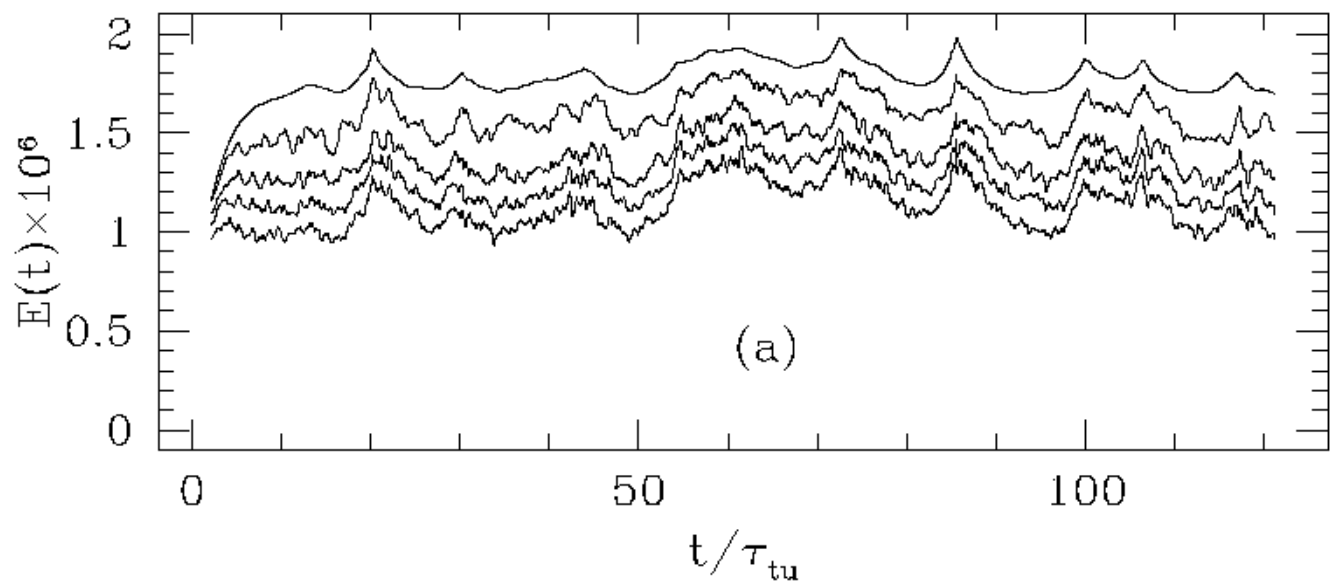
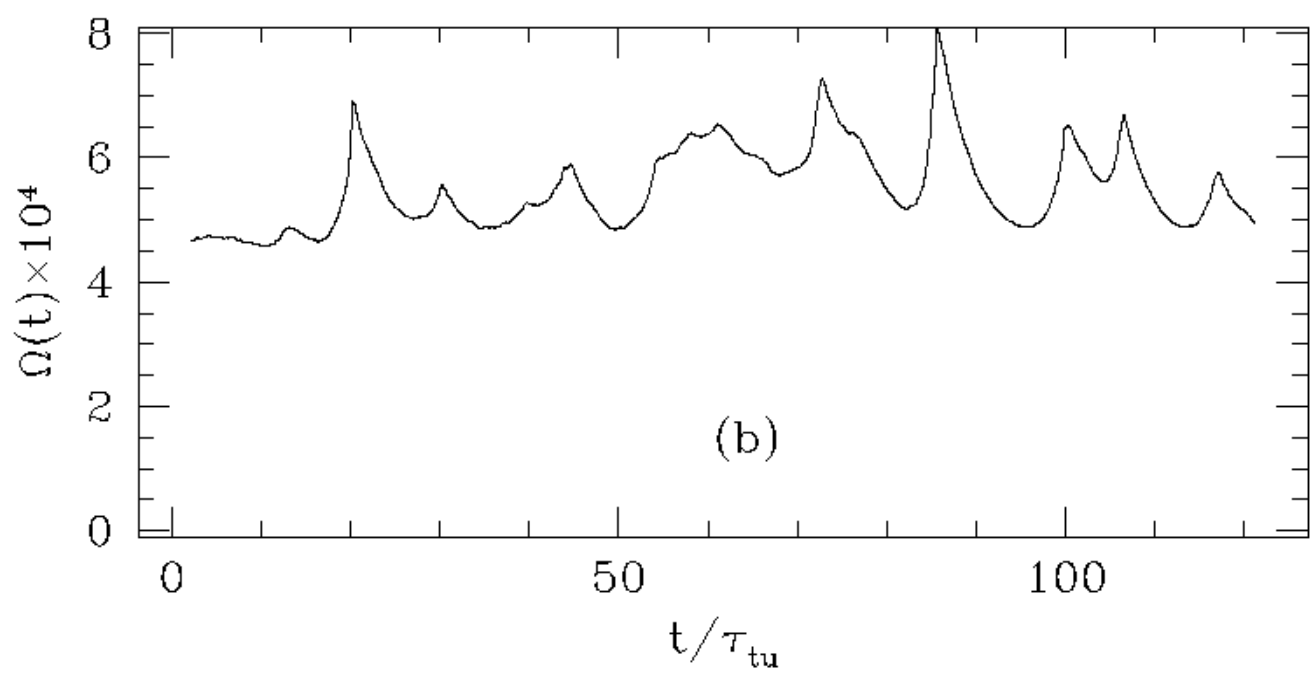
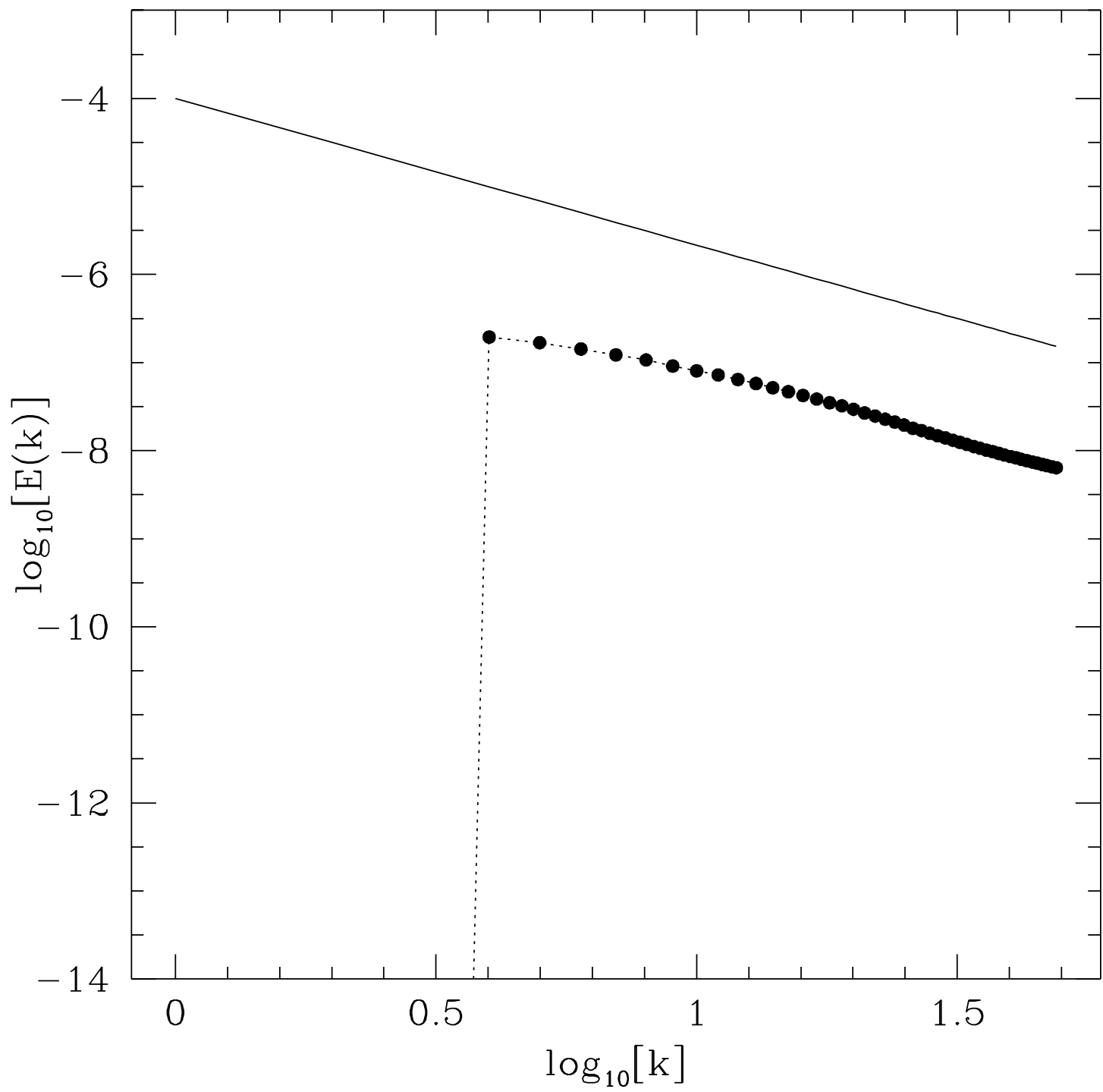
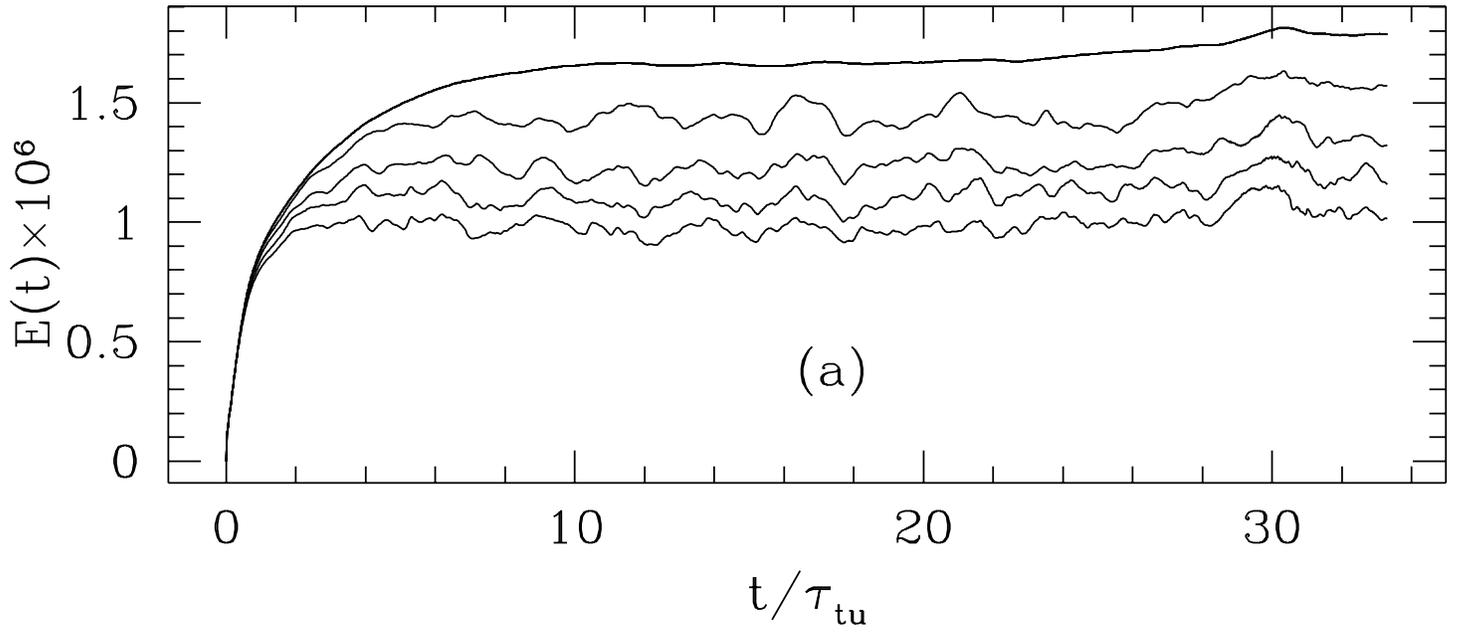
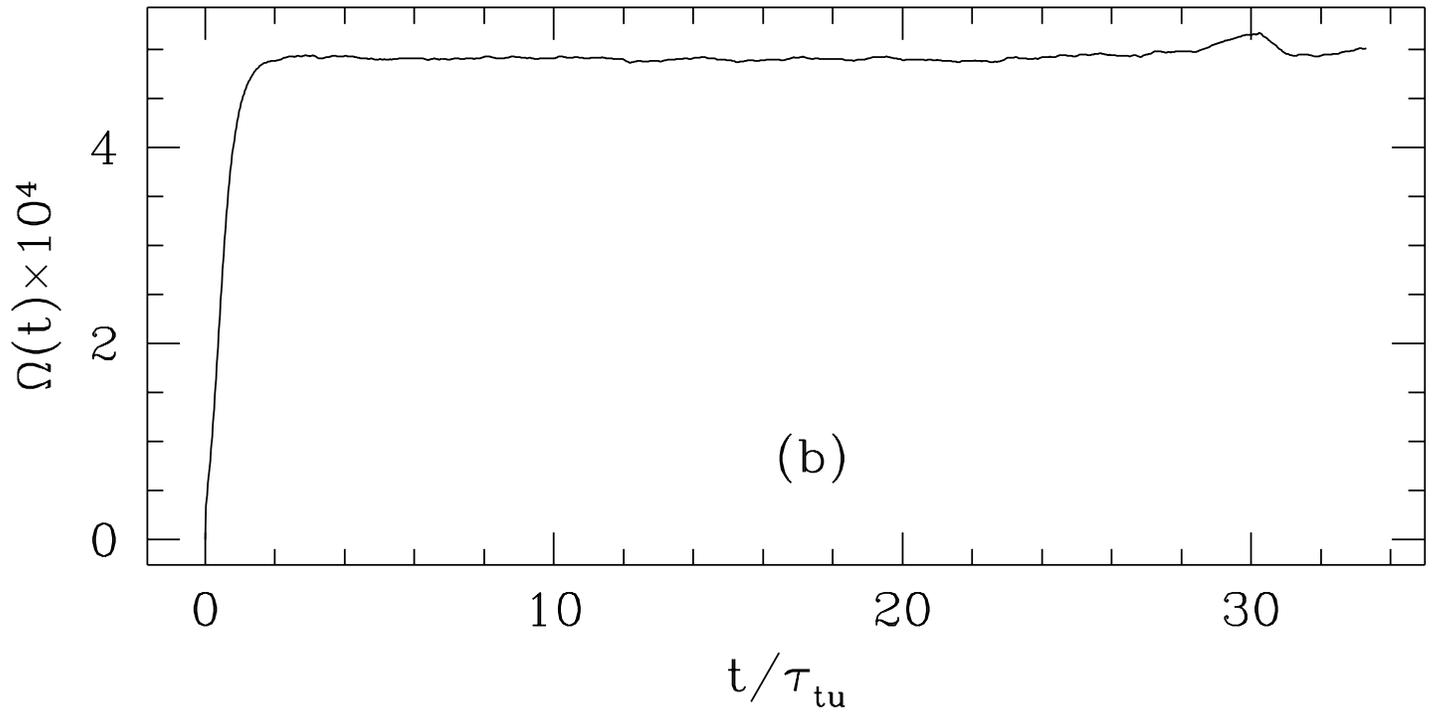
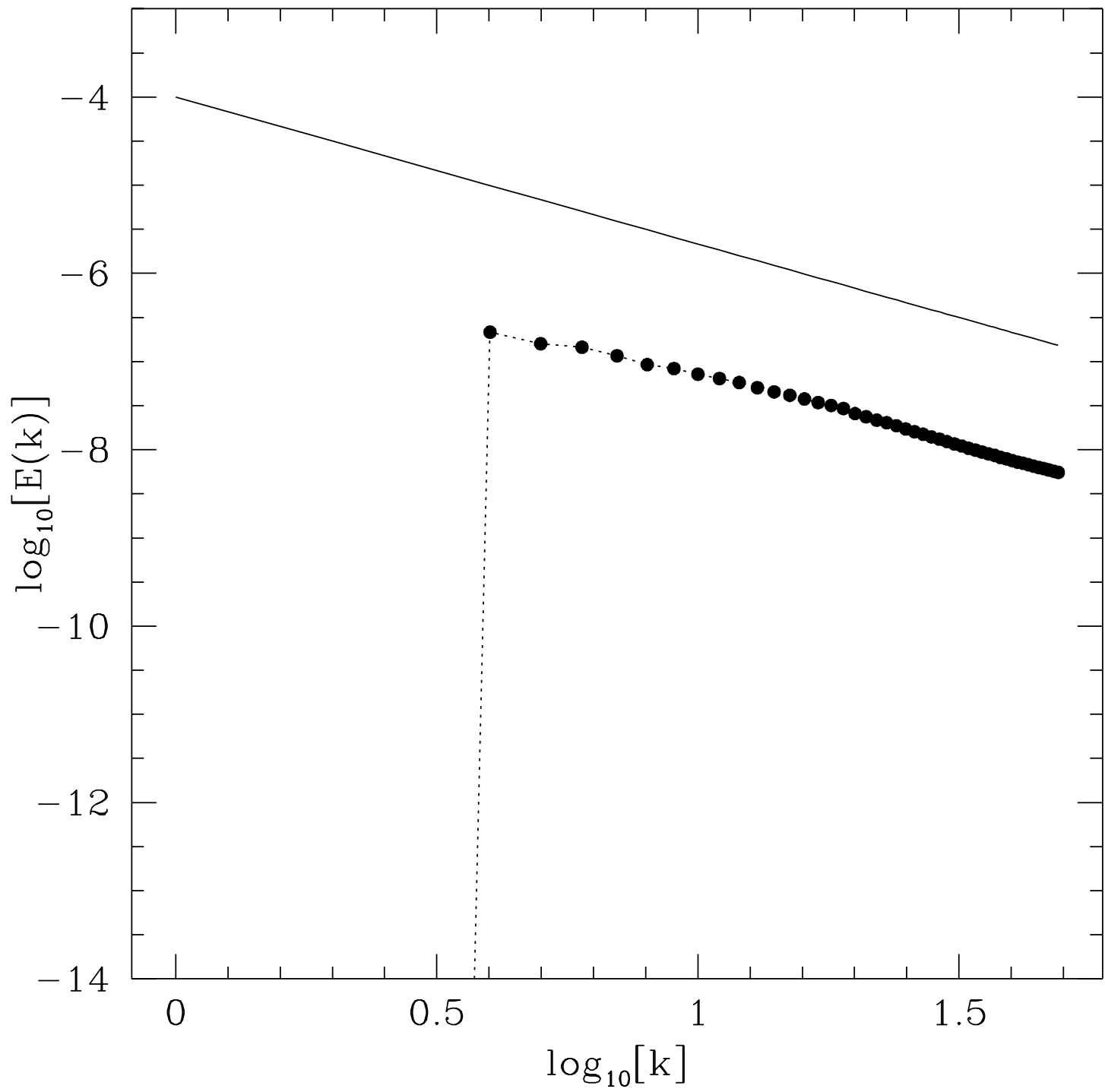
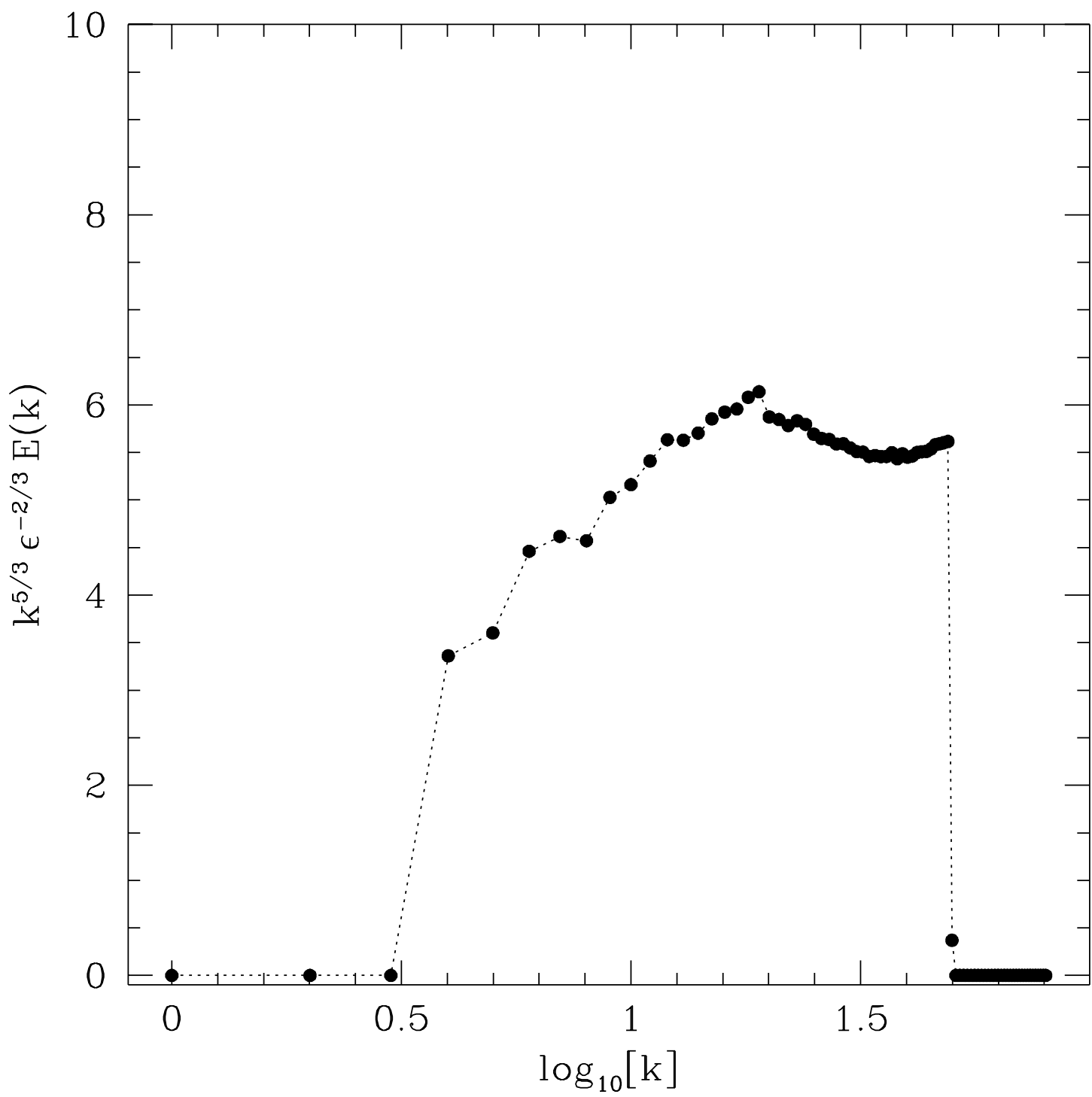
